\documentclass[structabstract]{aa}

\usepackage{txfonts}
\usepackage{graphicx}
\usepackage{natbib}
\bibpunct{(}{)}{;}{a}{}{,}

\begin{document}

  \title{A close look at the Centaurus A group of galaxies \\ 
II. Intermediate-age populations in early-type dwarfs\thanks{Based on observations collected at the European Southern Observatory, Paranal, Chile, within the observing Programme 073.B-0131.} \thanks{Table 3 is only available in electronic form at the CDS via anonymous ftp to cdsarc.u-strasbg.fr (130.79.128.5)
or via http://cdsweb.u-strasbg.fr/cgi-bin/qcat?J/A+A/ .}}
  \author{D. Crnojevi\'{c}\inst{1}\fnmsep\thanks{Member of IMPRS (International Max Planck Research School) for Astronomy \& Cosmic Physics at the University of Heidelberg and of the Heidelberg Graduate School for Fundamental Physics.} \and M. Rejkuba\inst{2} \and E. K. Grebel\inst{1} \and G. Da Costa\inst{3} \and H. Jerjen\inst{3}
}

 \institute{Astronomisches Rechen-Institut, Zentrum f\"{u}r Astronomie der
   Universit\"{a}t Heidelberg, M\"{o}nchhofstrasse 12-14, 69120 Heidelberg, Germany \newline
 \email{denija@ari.uni-heidelberg.de}
\and
 European Southern Observatory, Karl-Schwarzschild-Strasse 2, D-85748 Garching, Germany
\and
 Research School of Astronomy \& Astrophysics, Institute of Advanced Studies, Australian National University, Cotter Road, Weston Creek, ACT 2611, Australia
}

  \date{Received 19 July 2010 / Accepted 16 March 2011}

  \abstract
  {}{We investigate the resolved stellar content of early-type dwarf galaxies in the Centaurus A group, in order to estimate the fraction of their intermediate-age populations.}{We use near-infrared photometric data taken with the VLT/ISAAC instrument, together with previously analyzed archival HST/ACS data. The combination of the optical and infrared wavelength range permits us to firmly identify luminous asymptotic giant branch stars, which are indicative of an intermediate-age population in these galaxies.}{We consider one dwarf spheroidal (CenA-dE1) and two dwarf elliptical (SGC1319.1-4216 and ESO269-066) galaxies that are dominated by an old population. The most recent periods of star formation are estimated to have taken place between $\sim2$ and $\sim5$ Gyr ago for SGC1319.1-4216 and ESO269-066, and approximately 9 Gyr ago for CenA-dE1. For ESO269-066, we find that the intermediate-age populations are significantly more centrally concentrated than the predominantly old underlying stars. The intermediate-age population fraction is found to be low in the target galaxies, consistent with fractions of up to $\sim15\%$ of the total population. These values could be higher by a factor of two or three, if we consider the observational limitations and the recent discussion about the uncertainties in theoretical models. We suggest that there is a correlation between intermediate-age population fraction and proximity to the dominant group galaxy, with closer dwarfs having slightly smaller such fractions, although our sample is too small to draw firm conclusions.}{Even when considering our results as lower limits, the intermediate-age population fractions for the studied dwarfs are clearly much lower than those found in similar dwarfs around the Milky Way, but comparable to what is seen for the low-mass M31 companions. Our results confirm previous literature work by Rejkuba et al. (2006) about early-type dwarfs in the Centaurus A group.}

\titlerunning{A close look at the Centaurus A group of galaxies. II. Intermediate-age populations in early-type dwarfs}
  \maketitle

%________________________________________________________________

\section{Introduction}

The Centaurus A group of galaxies is, together with the more sparse Sculptor group, the nearest prominent galaxy group in the southern sky. In its center the peculiar giant elliptical galaxy NGC5128 (=CenA) is located, whose distance is 3.8 Mpc \citep{harrisg09}. Recent searches for new members of this group were published by \citet{cote97}, \citet{banks99}, and \citet{jerjen00b}. \citet{kara02} provided precise radial distance measurements for 17 dwarf members of the group using the tip of the red giant branch (TRGB) method. The current list of members of the CenA group contains 62 galaxies with radial velocities in the Local Group rest frame of $V_{LG}<550$ km s$^{-1}$ and angular distances from CenA of $< 30^\circ$ \citep{kara07}. 

Apart from the determinations of the individual distances, the study of the resolved stellar populations via deep imaging offers insights in the star formation history (SFH) of these systems. Over the last couple of decades, resolved stellar populations have been studied in most Local Group galaxies in great detail. The main result of all these studies is a surprizing variety of their SFHs \citep[see e.g.][]{grebel97, tolstoy09}. In order to explore the role of the environment in the evolution of these systems, it is desirable to investigate stellar populations and SFHs of dwarf galaxies in an environment that is different from the Local Group. With its higher density of galaxies, the Centaurus A group offers such a possibility.

In \citet{crnojevic10} we considered optical Hubble Space Telescope (HST) archival images to perform photometry of six early-type dwarfs in the Centaurus A group, for which the upper part of the red giant branch (RGB) can be resolved. We derived their photometric metallicity distribution functions and looked for stellar spatial gradients in their (predominantly) old populations. Similar work was done for early-type dwarfs in the M81 group by \citet{lianou10}. However, we pointed out that optical photometry alone is not enough to unambiguously estimate the fraction of intermediate-age populations (IAPs) in these objects, given the substantial amount of Galactic foreground contamination in the direction of the Centaurus A group. Previous studies have shown that luminous asymptotic giant branch (AGB) stars, indicative of such an IAP, are brighter in the near-infrared (NIR) bands, and that the foreground is easier to separate from the galactic stellar content with this kind of data \citep[e.g.,][]{rejkuba06, boyer09}. The combination of optical and NIR data is thus a powerful tool to investigate to which extent and at which ages the target dwarfs produced their IAPs.

Early-type dwarf galaxies of the Centaurus A group have been observed in NIR bands, with the Infrared Spectrometer And Array Camera (ISAAC) at European Southern Observatory Very Large Telescope (ESO VLT). The first results for two early-type dwarf members of this group were presented in \citet{rejkuba06}. We now further study three additional objects of the NIR sample.

The paper is organized as follows. We describe the data in \S \ref{data}, and present the derived color-magnitude diagrams (CMDs) for both wavelength sets in \S \ref{cmd_sec}. The IAPs are then investigated in detail in \S \ref{agb_sec}, and the discussion is presented in \S \ref{discuss}. Finally, we draw our conclusions in \S \ref{conclus}.

%________________________________________________________________

\section{Data and photometry} \label{data}

\begin{table*}
 \centering
\caption{Fundamental properties of the target dwarf galaxies.}
\label{infogen}
\begin{tabular}{lccccccccc}
\hline
\hline
Galaxy&RA&DEC&$T$&$D$&$D_{CenA}$&$E(B-V)$&$M_{V}$&$<$[Fe/H]$>_{med}$&Last SF\\
&(J2000)&(J2000)&&(Mpc)&(kpc)&&&(dex)&(Gyr)\\
\hline
\object{CenA-dE1, KK189}&$13\,12\,45.0$&$-41\,49\,55$&$-3$&$4.42\pm0.33$&$676\pm483$&$0.11$&$-11.99$&$-1.52\pm0.20$&$\sim9$\\
\object{SGC1319.1-4216, KK197}&$13\,22\,01.8$&$-42\,32\,08$&$-3$&$3.87\pm0.27$&$198\pm112$&$0.15$&$-13.04$&$-1.08\pm0.41$&$2-5$\\
\object{ESO269-066, KK190}&$13\,13\,09.2$&$-44\,53\,24$&$-5$&$3.82\pm0.26$&$113\pm412$&$0.09$&$-13.89$&$-1.21\pm0.33$&$2-3$\\
\hline
\end{tabular}
\tablefoot{
The columns are as follows: column (1): name of the galaxy; (2-3): equatorial coordinates (J2000, units of right ascension are hours, minutes, and seconds, and units of declination are degrees, arcminutes, and arcseconds); (4): de Vaucouleurs' morphological T-type from \citet{kara05}; (5): distance of the galaxy derived by \citet{kara07} with the TRGB method; (6): deprojected distance of the galaxy from the dominant elliptical CenA \citep{crnojevic10}; (7) foreground reddening from \citet{schlegel98}; (8): total absolute $V$ magnitude from \citet{georgiev08}; (9): photometrically derived median metallicity and intrinsic metallicity dispersion of RGB stars, taken from \citet{crnojevic10}; and (10): epoch of most recent significant star formation episode, as derived in Sect. \ref{age_sec}.
}
\end{table*}

Within the ESO observing programme 073.B-0131 we collected NIR observations of 14 early-type dwarfs in the Centaurus A Group. We choose to study in detail here those galaxies that could be fully resolved (due to good seeing conditions, i.e. seeing $<0.6$ arcsec) in their stellar content from these observations, and for which archival HST data were already analyzed by \citet{crnojevic10}, similarly to what was already done by \citet{rejkuba06}.

The sample consists of three objects, namely CenA-dE1, SGC1319.1-4216 and ESO269-066 (in order of increasing luminosity). We report the main properties of the target galaxies in Table \ref{infogen}. 

We will now separately consider the optical and NIR data for the studied galaxies.

\subsection{Optical photometry} \label{dataopt}

\begin{figure}
 \centering
  \includegraphics[width=7cm]{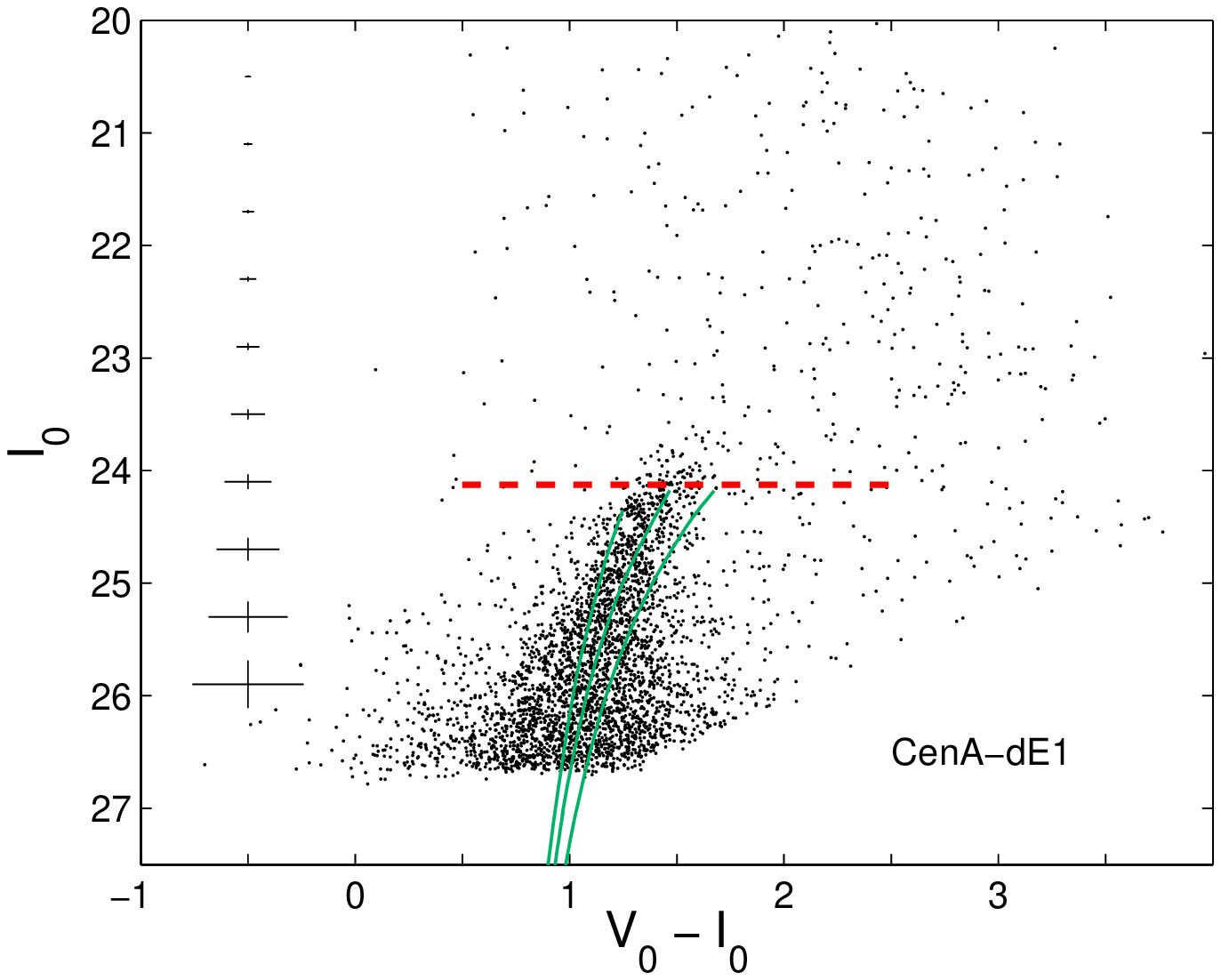}
  \includegraphics[width=7cm]{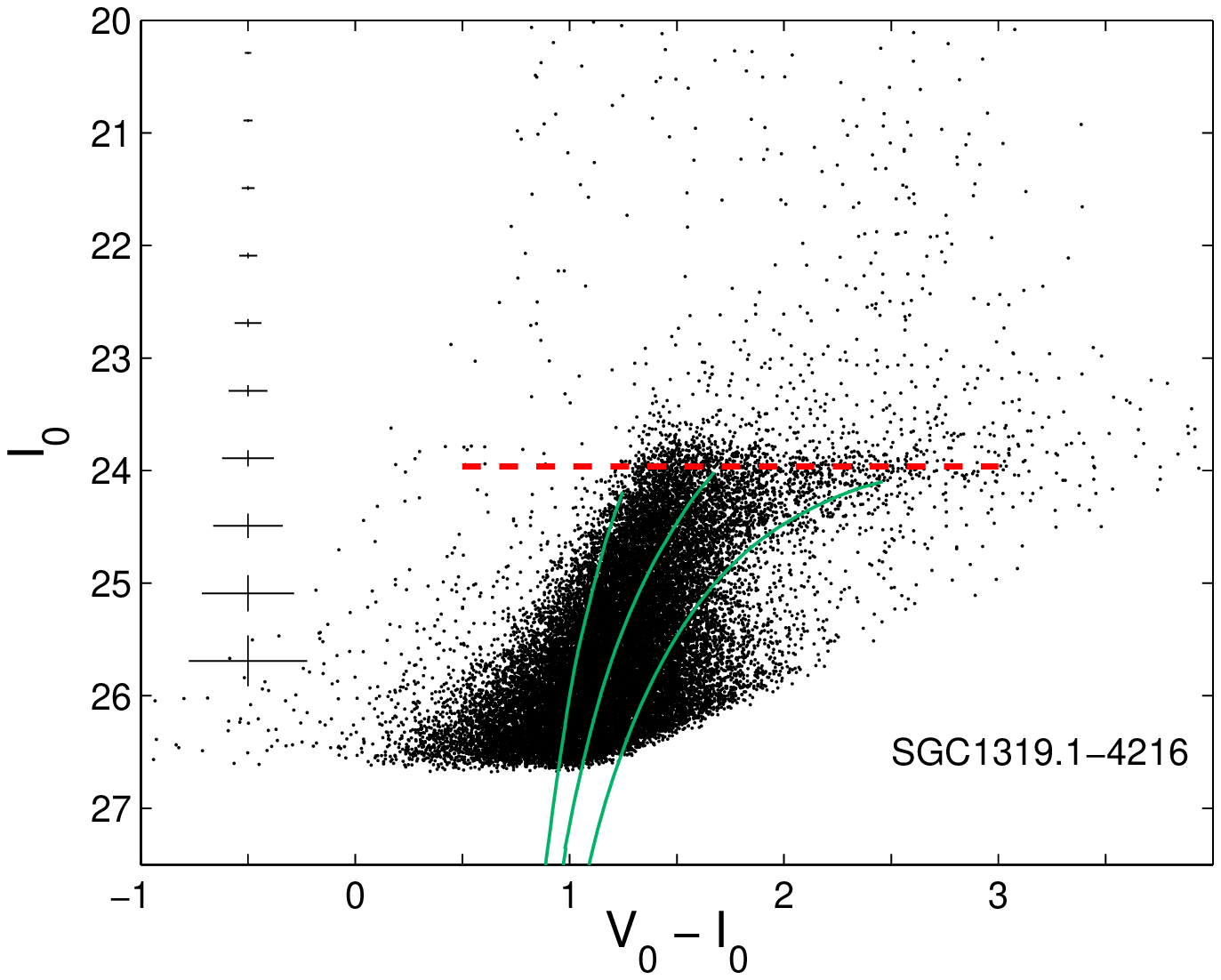}
  \includegraphics[width=7cm]{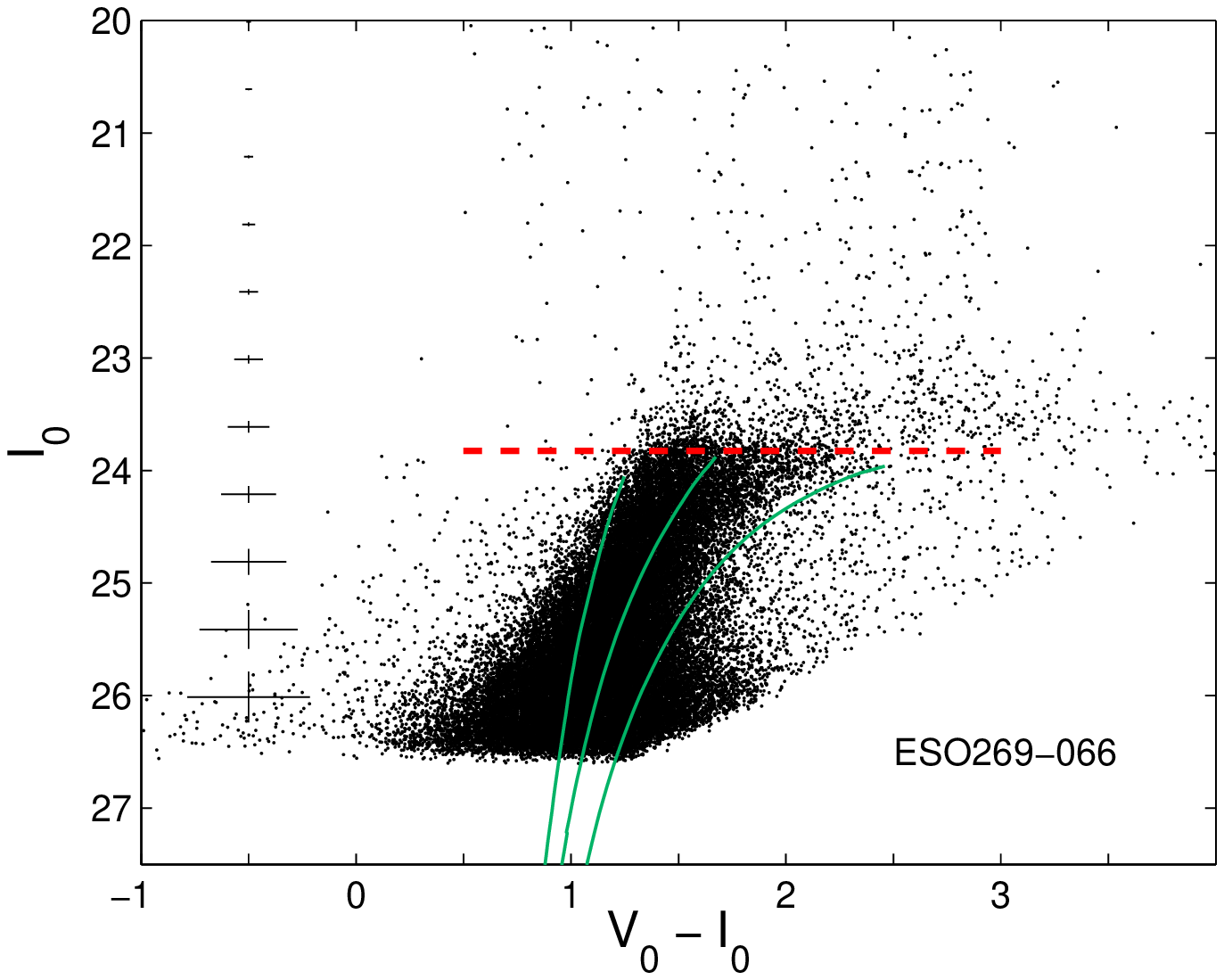}
 \caption{\footnotesize{Optical (dereddened) CMDs of the target galaxies, ordered by increasing luminosity. The dominant feature of the CMDs is a prominent RGB, with a small presence of luminous AGB stars above the TRGB (indicated by a red dashed line). Representative photometric errorbars are plotted on the left side of the CMDs. Green lines are stellar isochrones with a fixed age of 10 Gyr, and varying metallicity: [Fe/H]$=-2.5$, $-1.52$ and $-1.1$ (from the blue to the red side) for CenA-dE1; [Fe/H]$=-2.5$, $-1.08$ and $-0.4$ for SGC1319.1-4216; [Fe/H]$=-2.5$, $-1.21$ and $-0.4$ for ESO269-066 (see text for details).}}
 \label{cmdopt}
\end{figure}

The archival optical HST data, taken with the Advanced Camera for Surveys (ACS), were already presented and discussed in \citet{crnojevic10}, and are not described in detail here again. For each one of the target galaxies there is one 1200 seconds exposure in the $F606W$ filter (roughly corresponding to the $V$-band) and one 900 seconds exposure in the $F814W$ filter (corresponding to the $I$-band). The photometry was performed using the DOLPHOT package \citep{dolphin02}, which was also used to estimate photometric errors and observational incompleteness via artificial star tests. Typical photometric errorbars from artificial star tests are shown in the CMDs in Fig. \ref{cmdopt}. The DOLPHOT package additionally provides magnitudes in the ground-based $V$- and $I$- bands, which are the ones we use throughout the paper. The $50\%$ completeness level in the $V$-band lies in the range $\sim27.2-26.8$, where the first value refers to the least crowded of the galaxies in this sample, CenA-dE1, and the second to the most crowded, ESO269-066. In the $I$-band, the $50\%$ completeness ranges from $\sim26.6$ to $\sim26.1$. The number of stars in the final photometric catalogs (after quality cuts) is 3675, 26713 and 39064 for CenA-dE1, SGC1319.1-4216 and ESO269-066, respectively.

\subsection{Near-infrared photometry} \label{datanir}

\begin{table}
 \centering
\caption{NIR imaging observing log.}
\label{infonir}
\begin{tabular}{lccccc}
\hline
\hline
Galaxy&Date&F&t$_{exp}$&AM&Seeing\\
&(dd/mm/yy)&&(sec)&&(arcsec)\\
\hline
\object{CenA-dE1}&21/04/04&$K_S$&2352&1.23&0.59\\
&11/05/04&$K_S$&2352&1.08&0.34\\
&27/05/04&$K_S$&2352&1.21&0.41\\
&27/05/04&$J_S$&2100&1.11&0.44\\
\object{SGC1319.1-4216}&14/05/04&$K_S$&2352&1.44&0.43\\
&10/07/04&$K_S$&2352&1.11&0.47\\
&10/07/04&$J_S$&2100&1.20&0.77\\
&19/07/04&$J_S$&2100&1.32&0.44\\
\object{ESO269-066}&11/05/04&$K_S$&2352&1.20&0.41\\
&31/05/04&$K_S$&2352&1.12&0.34\\
&31/05/04&$J_S$&2100&1.21&0.41\\
\hline
\end{tabular}
\tablefoot{
The columns contain: (1): galaxy name; (2): date of observation; (3): filter; (4): exposure time (for the dithered sequences, units of sec); (5): airmass (given for the central image in each sequence); and (6): seeing of the combined images (as measured on the combined images, units of arcsec).
}
\end{table}

Deep NIR images of the target dwarfs were taken in service mode with the short wavelength arm of ISAAC NIR array at the VLT at ESO Paranal Observatory. The field of view of the short wavelength arm of ISAAC is $2.5\times2.5$ arcmin$^2$ and the detector has a pixel scale of $0.148$ arcsec. Each galaxy was observed once in the $J_S$-band by coadding a sequence of dithered short exposures, amounting to a total exposure time of 2100 sec per galaxy. $K_S$-band images were taken with a similar strategy at two different epochs, amounting to a total exposure time of 4704 sec per galaxy. The observing log is reported in Tab. \ref{infonir}.

The standard procedure in reducing IR data was used, and the details of ISAAC data reduction with IRAF\footnote{IRAF is supported by the National Optical Astronomy Observatories, which are operated by the Association of Universities for Research in Astronomy, Inc., under cooperative agreement with the National Science Foundation.} are described by \citet{rejkuba01}. At the end of the reduction all the images taken in a single dithered sequence were combined. From now on when we refer to an image observed in the $J_S$-band or $K_S$-band we always refer to these combined dithered sequences, and we also drop the ``s'' subscript and just use the nomenclature convention $J$- and $K$-band.

For all the targets point spread function (PSF) fitting photometry was performed using the suite of DAOPHOT and ALLFRAME programs \citep{stetson87, stetson94}. The procedure for each galaxy target included the following steps. For each image we detected all the point sources and determined their PSF using at least 30 relatively bright, non-saturated stars, well spread across the field. For the observations in both filters the coordinate transformation were derived. The complete star list was then created from the median combined image. In case of large variations of seeing between different images, the worst seeing image was not used in the median combination. PSF fitting photometry using this star list and coordinate transformations was performed simultaneously on each image using ALLFRAME \citep{stetson94}. The final photometric catalog for each galaxy contains all the sources that could be measured in both the $J$-band and at least one $K$-band image. We further apply the following quality cuts to the catalog: the photometric errors (as measured by ALLFRAME) need to be smaller than 0.3 mag in both bands; the sharpness parameter has an absolute value $\le2$; and we impose $\chi\le1.5$. The number of stars from the NIR photometry in each galaxy is: 347 (CenA-dE1), 1505 (SGC1319.1-4216) and 3428 (ESO269-066).

We have tied our photometry to the 2MASS photometric system \citep{carpenter01}\footnote{http://www.ipac.caltech.edu/2mass/releases/allsky/doc/sec6\_4b.html/.} by matching all the point sources from 2MASS observed in our fields with our $J$-band and $K$-band detections (the adopted photometric system is the Vega system). Zero points for each galaxy included thus the correction for the atmospheric extinction. Typically more than 7 stars had 2MASS magnitudes, resulting in zeropoint uncertainties of the order of 0.02 to 0.1 mag in the $J$-band and 0.03 to 0.15 mag in the $K$-band.

\begin{figure}
 \centering
  \includegraphics[width=7cm]{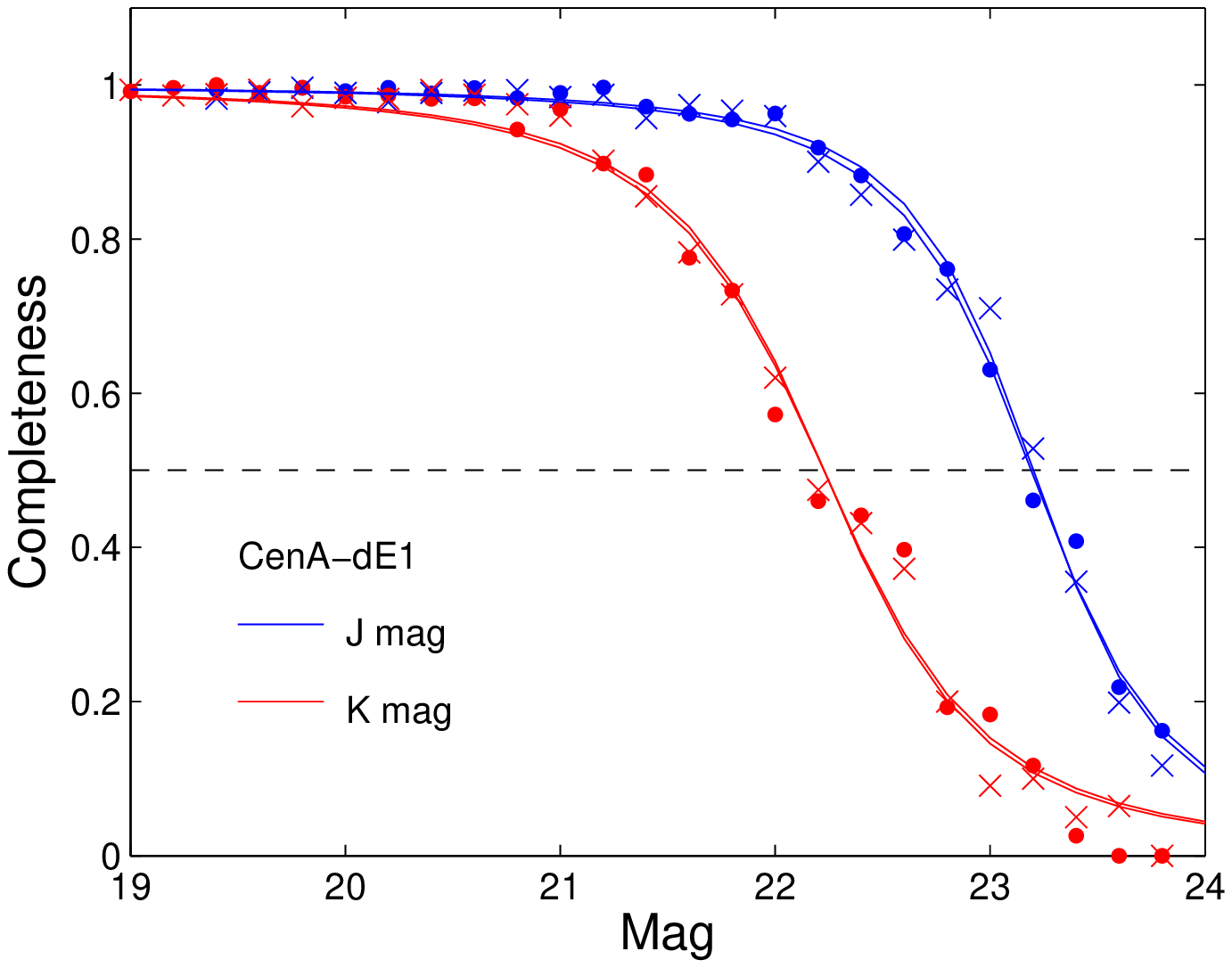}
  \includegraphics[width=7cm]{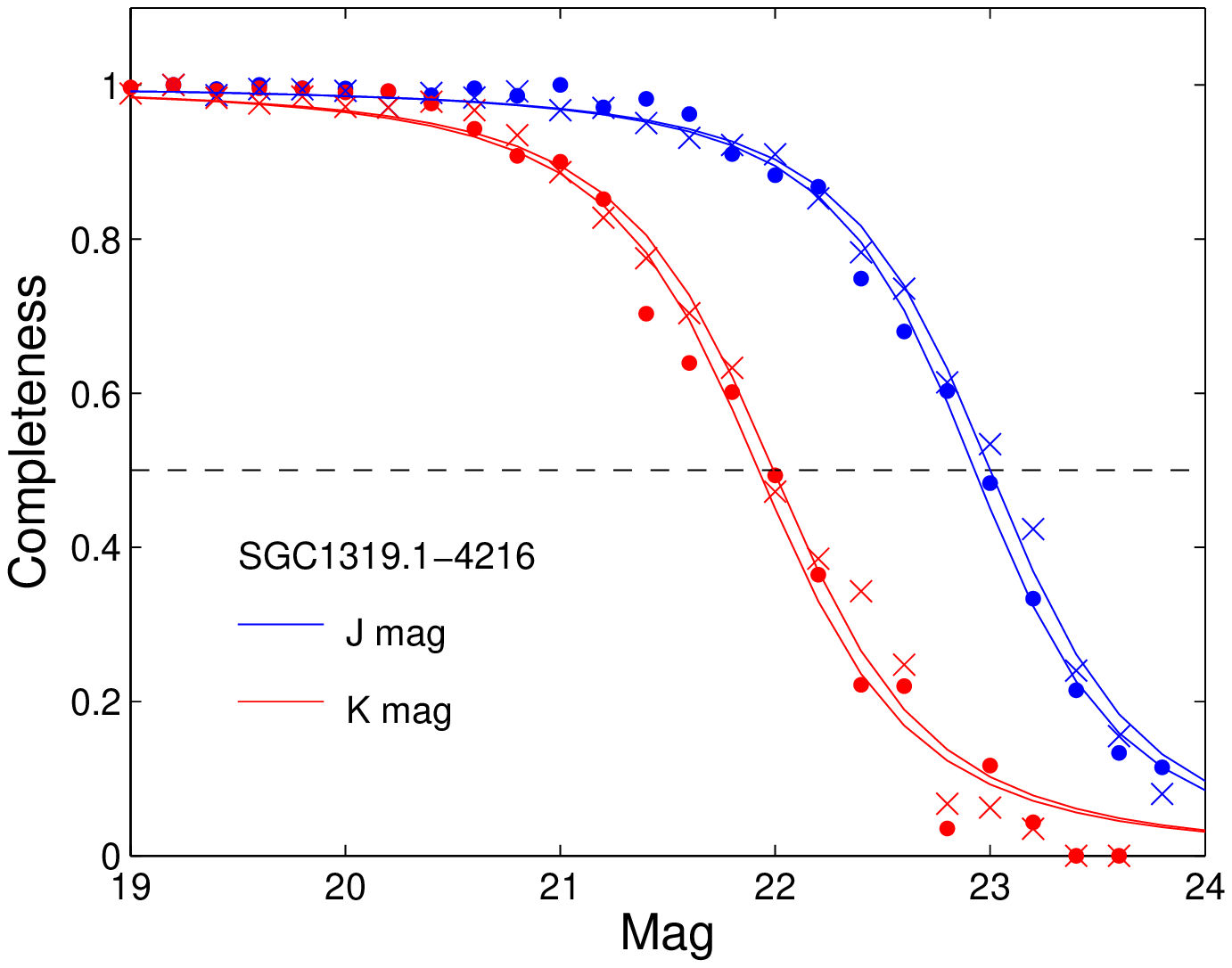}
  \includegraphics[width=7cm]{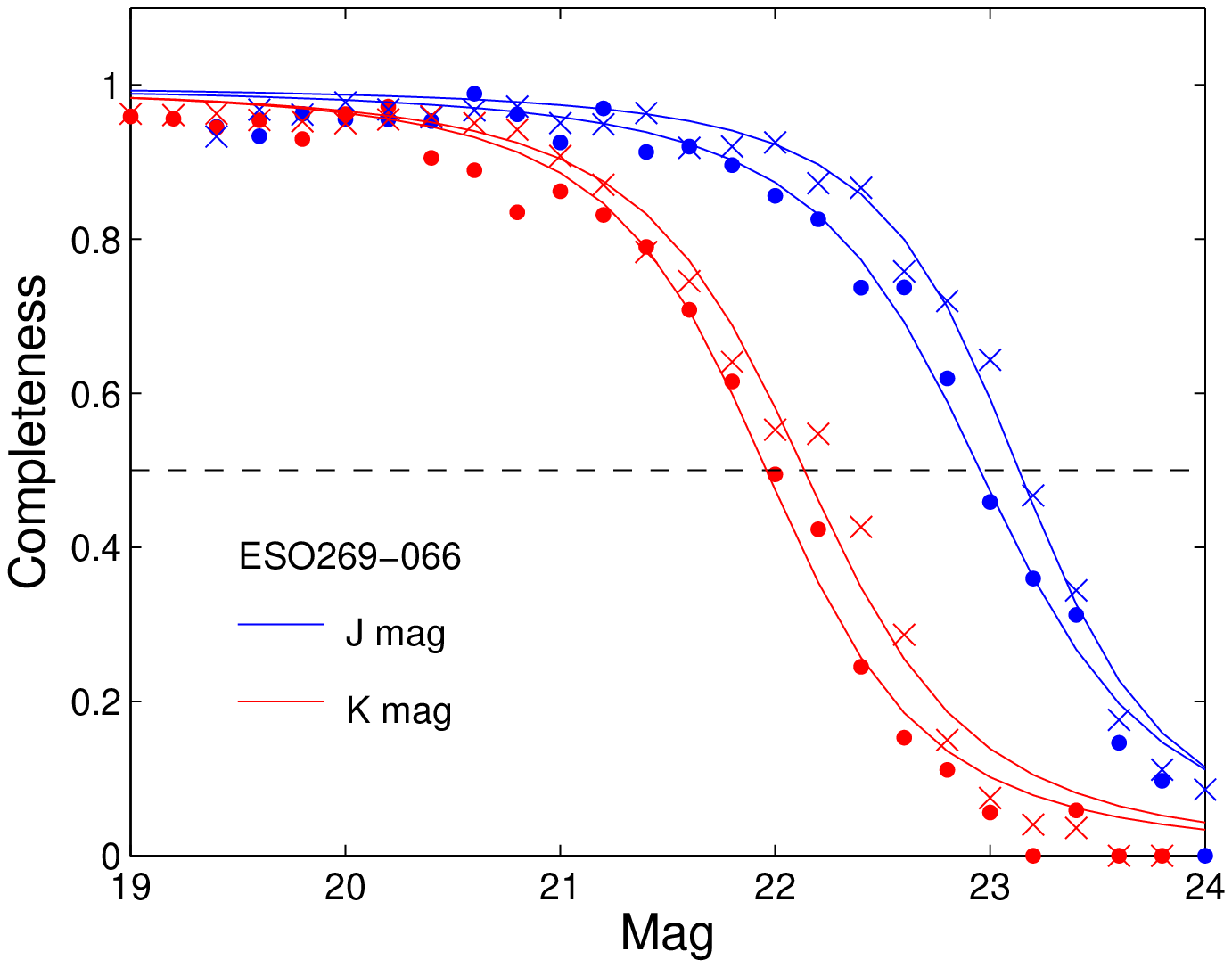}
 \caption{\footnotesize{$J$- and $K$-band completeness curves for the target galaxies. The dashed line shows the $50\%$ completeness level. Dots delineate curves that were computed using stars from the central region of the galaxy, while crosses are for stars located in the outskirts of the galaxy (see text for details). We overplot the best-fitting analytic completeness functions as solid lines \citep{fleming95}. Even for the most crowded field (ESO269-066), the inner and outer completeness curves differ very little.}}
 \label{compl}
\end{figure}

Completeness and magnitude errors were measured using artificial star tests. Fake stars were added to each image using its measured PSF and adding the expected noise. Photometry of all the stars was then performed and typical photometric errors were derived as the difference between input and recovered magnitude. We also compute the percentage of recovered simulated stars, and test to which extent stellar crowding affects the completeness of the observations. CenA-dE1 is the least crowded galaxy of our sample, while ESO269-066 is the most crowded. We choose an elliptical radius that divides the stellar sample into a central subsample (which should be the most affected by crowding) and a subsample containing stars in the outskirts of the galaxies. To define this radius, we use the stellar density profiles computed in \citet{crnojevic10}. CenA-dE1 is off-center in the ISAAC field of view, so we can use the galaxy's limiting (=tidal) radius, beyond which almost only field stars are found. SGC1319.1-4216 and ESO296-066 are more extended than the field of view, so we adopt the half-light radius from \citet{crnojevic10}. The completeness curves are shown in Fig. \ref{compl}, for both the $J$- and $K$-band and for both the central and external stellar subsamples. We fit the curves with the analytic function introduced by \citet{fleming95}
\begin{displaymath}
f=\frac{1}{2}\left[1-\frac{\alpha(m-m_0)}{\sqrt{1+\alpha^2(m-m_0)^2}}\right],
\end{displaymath}
and overplot the best-fit solutions in Fig. \ref{compl}. We note that completeness curves for particularly crowded fields can be also well fitted by alternative analytical functions \citep[e.g., the skewed Fermi law, see][]{puzia99}, but the choice of the fitting function does not affect our conclusions. Also in the case where the crowding is higher, namely for ESO269-066, the completeness is not changing significantly as a function of galactic radius. For example, for ESO269-066 the $50\%$ completeness limit (in both bands) is only $\sim0.2$ mag fainter for the external subsample than for the central one. From now on we will thus assume the completeness to be constant with radius. The photometric errors we derive are shown for each galaxy in the Figures of the next Section as representative errorbars in the CMDs (they do not include zero point uncertainties).

%________________________________________________________________

\section{Color-magnitude diagrams} \label{cmd_sec}

\subsection{Optical CMDs} \label{cmdopt_sec}

\begin{figure}
 \centering
  \includegraphics[width=9cm]{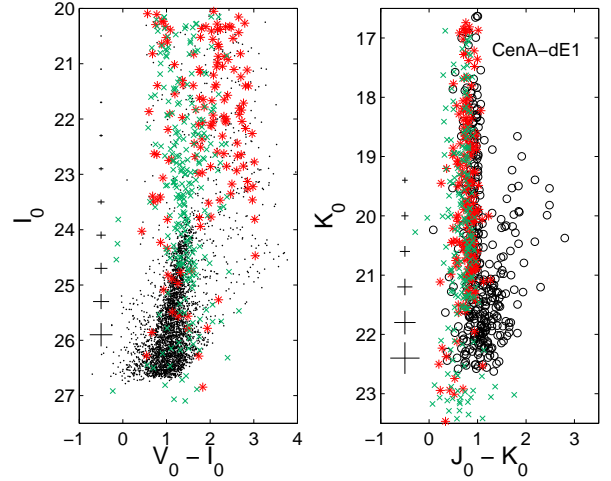}
 \caption{\footnotesize{Optical (\emph{left panel}) and NIR (\emph{right panel}) dereddened CMDs for CenA-dE1. Representative photometric errorbars are plotted on the left side of the CMDs. The green crosses represent the Galactic foreground contamination as predicted by TRILEGAL models, while the red asterisks are simulated using the Besan\c{c}on models. The simulated data include corrections for photometric errors and account for incompleteness effects. The models show that the optical CMD is heavily contaminated by foreground objects in the region above the TRGB, while in the NIR CMD is it easier to disentangle foreground stars from the stars belonging to CenA-dE1.}}
 \label{cont}
\end{figure}

The optical CMDs of the target galaxies are shown in Fig. \ref{cmdopt}. The data were dereddened taking the Galactic foreground reddening values from the NASA/IPAC Extragalactic Database (NED), which are based on the Schlegel extinction maps \citep{schlegel98}. It was assumed that these galaxies do not have a significant internal extinction due to the lack of current star formation. While locally enhanced extinction values could be present (i.e., around dust obscured carbon-rich AGB stars), this will only affect the individual star itself. We discuss in Sect. \ref{agb_sec} the possibility of missing such stars in our sample.

It can be seen that the stellar populations are dominated by a prominent RGB, which is indicative of an old population, but an accurate age-dating is not possible due to the metallicity-age degeneracy present in the RGB. The observations reach $\sim2.5$ mag ($I$-band) below the tip of the RGB. The latter is computed from the $I$-band luminosity function using a Sobel edge-detection filter \citep{lee93}. We find $I_{0,TRGB}=24.13\pm0.12$, $23.96\pm0.13$ and $23.83\pm0.12$ for CenA-dE1, SGC1319.1-4216 and ESO269-066, respectively. Not surprisingly, our results agree with the results of \citet{kara07} derived using the same dataset ($I_{0,TRGB}=24.18\pm0.07$, $23.89\pm0.04$ and $23.86\pm0.04$, respectively).

For the RGB stars, median metallicities and internal metallicity dispersions were derived by \citet{crnojevic10} assuming that they are mainly old stars ($\gtrsim10$ Gyr). Stellar isochrones from the Dartmouth evolutionary database \citep{dotter08} were used. First, a fixed age of 10 Gyr was assumed, and the metallicity was varied. Then, a metallicity value for each RGB star located between the TRGB and $\sim1$ mag below it was interpolated from this grid of isochrones, and a metallicity distribution function derived for the galaxy. In Fig. \ref{cmdopt}, we overplot three isochrones on each CMD: the first has the lowest metallicity value for that galaxy (corresponding to the lowest metallicity available from the isochrone set), the second represents the median metallicity (also reported in Tab. \ref{infogen}), and the last the highest metallicity value found.

With the current observations, we can exclude the presence of stars younger than $\sim500$ Myr in these galaxies (from the absence of an upper main sequence or supergiant stars), but above the TRGB of each galaxy we can see a number of stars that are probably luminous AGB stars. They are the bright tip of the iceberg of an IAP ($\sim1$ to 9 Gyr), which we expect to find at this position for metal-poor stellar populations. For populations more metal-rich than [Fe/H]$\sim-1.0$, some old and metal-rich stars may also be found above the TRGB, but our target objects are predominantly metal-poor (see Tab. \ref{infogen}), so we conclude that the presence of such stars is not significant in our sample.

Unfortunately, the low Galactic latitude of the Centaurus A group ($b\sim20^{\circ}$) means that a substantial amount of Galactic foreground stars contaminate the CMDs of our target galaxies. We simulate the expected foreground using both the TRILEGAL models \citep{girardi05} and the Besan\c{c}on models \citep{robin03}. An example of the expected contamination in our CMDs is shown in Fig. \ref{cont}. We plot both optical and NIR CMDs for CenA-dE1, and overplot the positions of foreground stars predicted by the two models (blue crosses for TRILEGAL and red asterisks for Besan\c{c}on). We convolve the simulated data with photometric errors and take into account incompleteness effects, and we stress that the simulated Galactic stars are only one random realization of the adopted models. Just for comparison purposes, we only plot stars simulated by Besan\c{c}on models that have masses $>0.15$M$_{\odot}$, since the TRILEGAL models do not include them in their computation. These excluded few, very low-mass stars would have the effect of extending the Besan\c{c}on plotted sequence to slightly redder colors. We can notice that at magnitudes $K_0\lesssim20.5$ and colors $J_0-K_0\lesssim1$ (where only foreground stars are found), the number counts resulting from the models are comparable to each other, and similar to the number of observed stars. We stress, however, that the models give slightly different results, and in particular the Besan\c{c}on model reaches slightly redder colors, but a comparison between the models is beyond the goals of this study. Overall, it is clear from the left panel that the luminous AGB region is the most affected in the optical, and we have no way of determining which stars belong to the dwarf galaxy and which are part of the foreground. We will come back to the NIR CMD in the following Section. In \citet{crnojevic10} we gave a rough estimate of the possible fraction of the IAP for the target early-type dwarfs by using the fuel consumption theorem (originally introduced by \citealt{renzini86}, see also \citealt{armand93}). We point out that the IAP fractions we report throughout this work are intended as numbers of stars (i.e., number of luminous AGB stars to old RGB stars). The results were the following: for CenA-dE1 and SGC1319.1-4216 the IAP fraction is $\sim10\%$, while for ESO269-066 it is $\sim15\%$ of the entire population.

\subsection{Near-infrared CMDs} \label{cmdnir_sec}

\begin{figure}
 \centering
  \includegraphics[width=9cm]{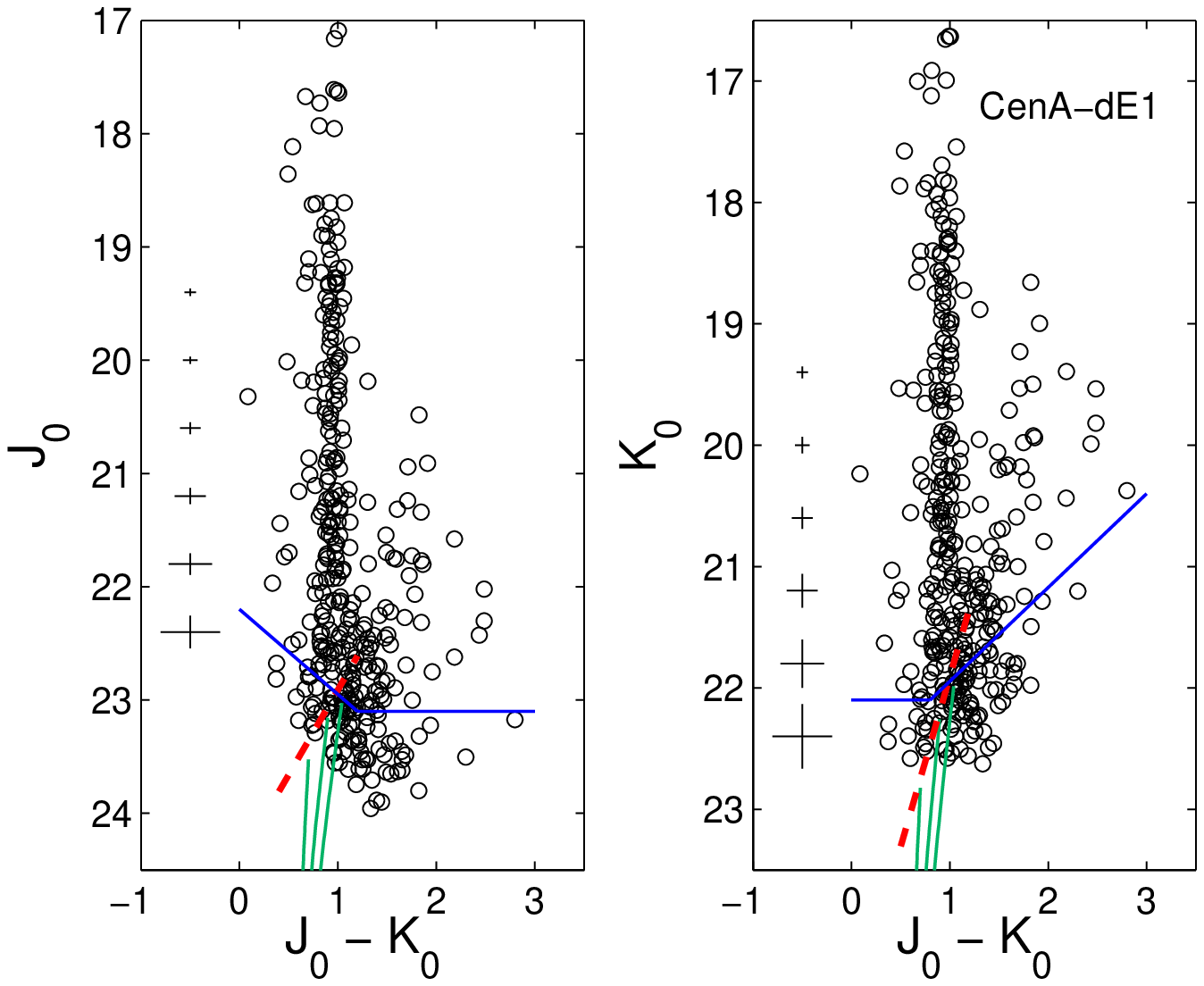}
  \includegraphics[width=9cm]{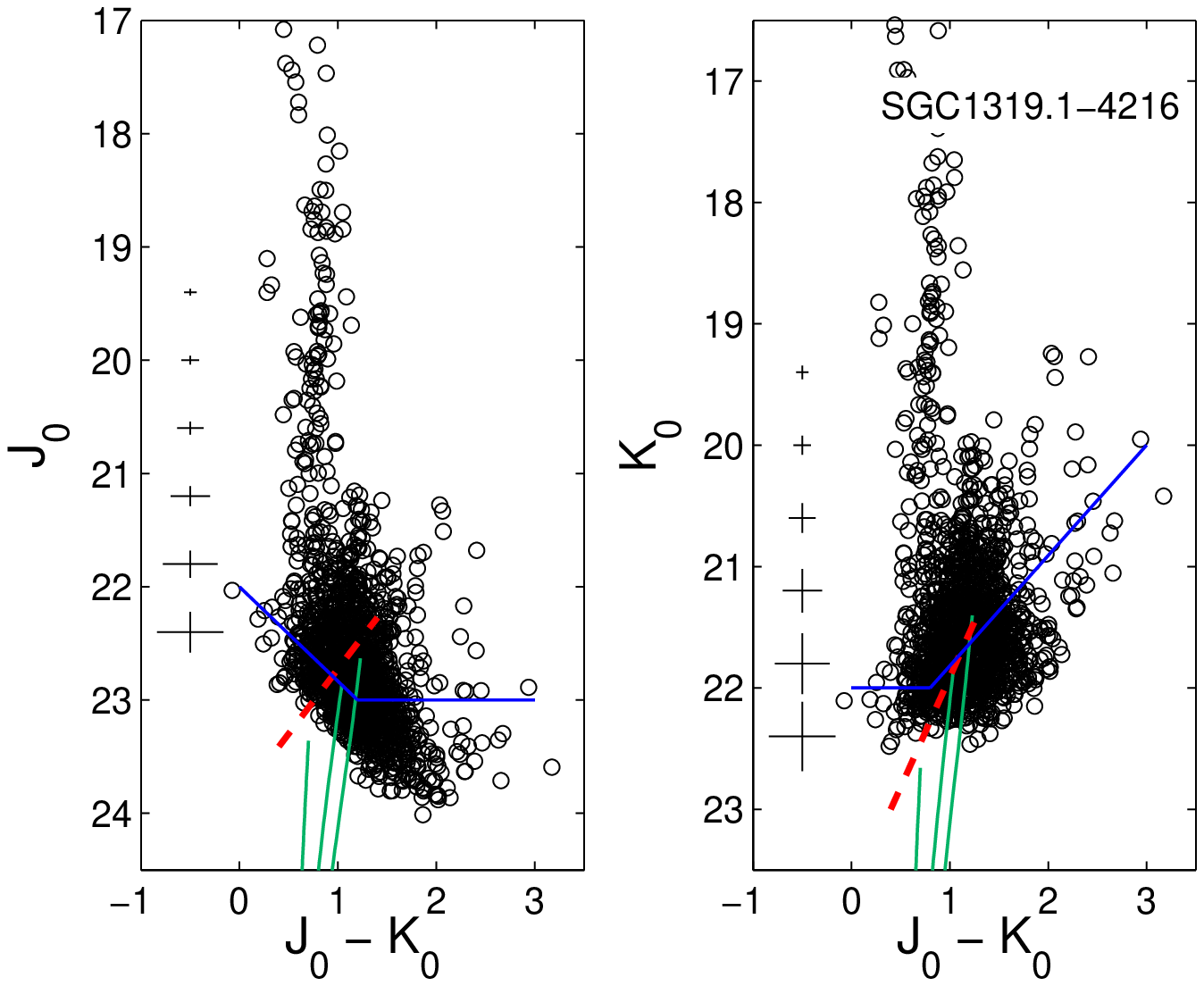}
  \includegraphics[width=9cm]{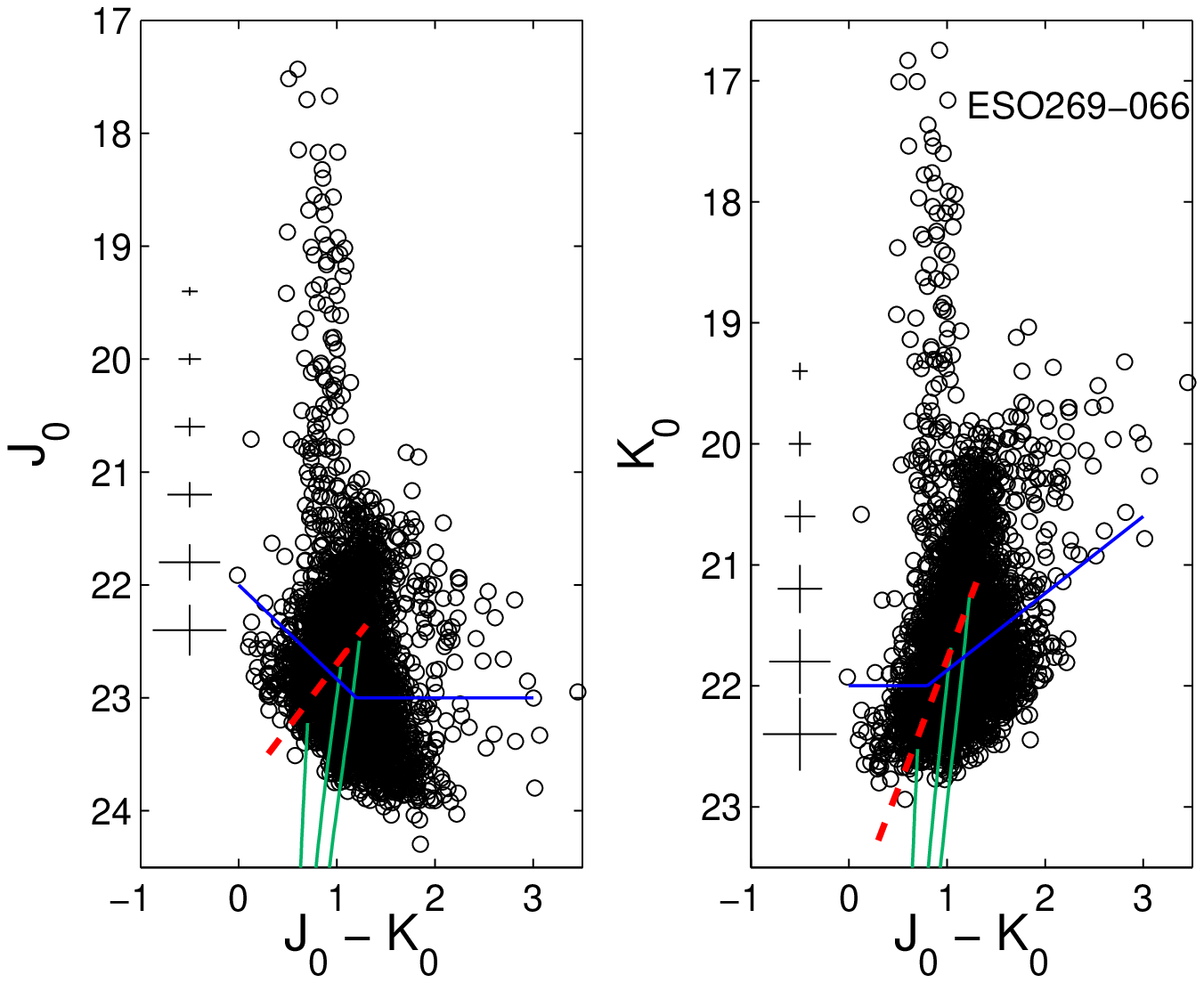}
 \caption{\footnotesize{NIR (dereddened) CMDs of the target galaxies, ordered by increasing luminosity. Representative photometric errorbars are plotted along the CMDs. The red dashed lines indicate the expected position of the TRGB as a function of metallicity (see text for details). The green lines are isochrones with a fixed age of 10 Gyr and spanning the metallicity range of each target galaxy, with the same values as in Fig. \ref{cmdopt}. The blue solid lines indicate the $50\%$ completeness limits.\vspace{1cm}}}
 \label{cmdnir}
\end{figure}

\begin{figure}
 \centering
  \includegraphics[width=7cm]{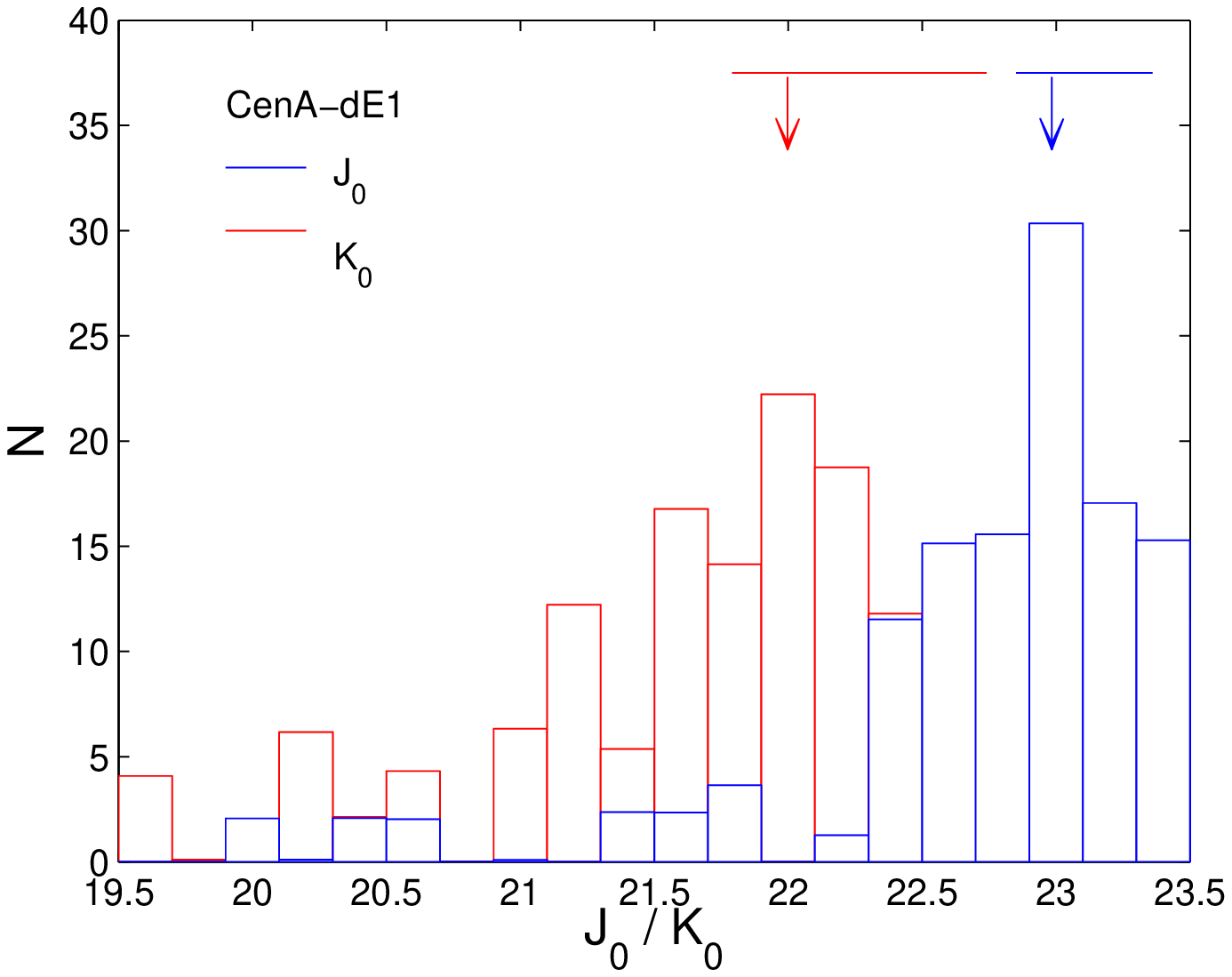}
  \includegraphics[width=7cm]{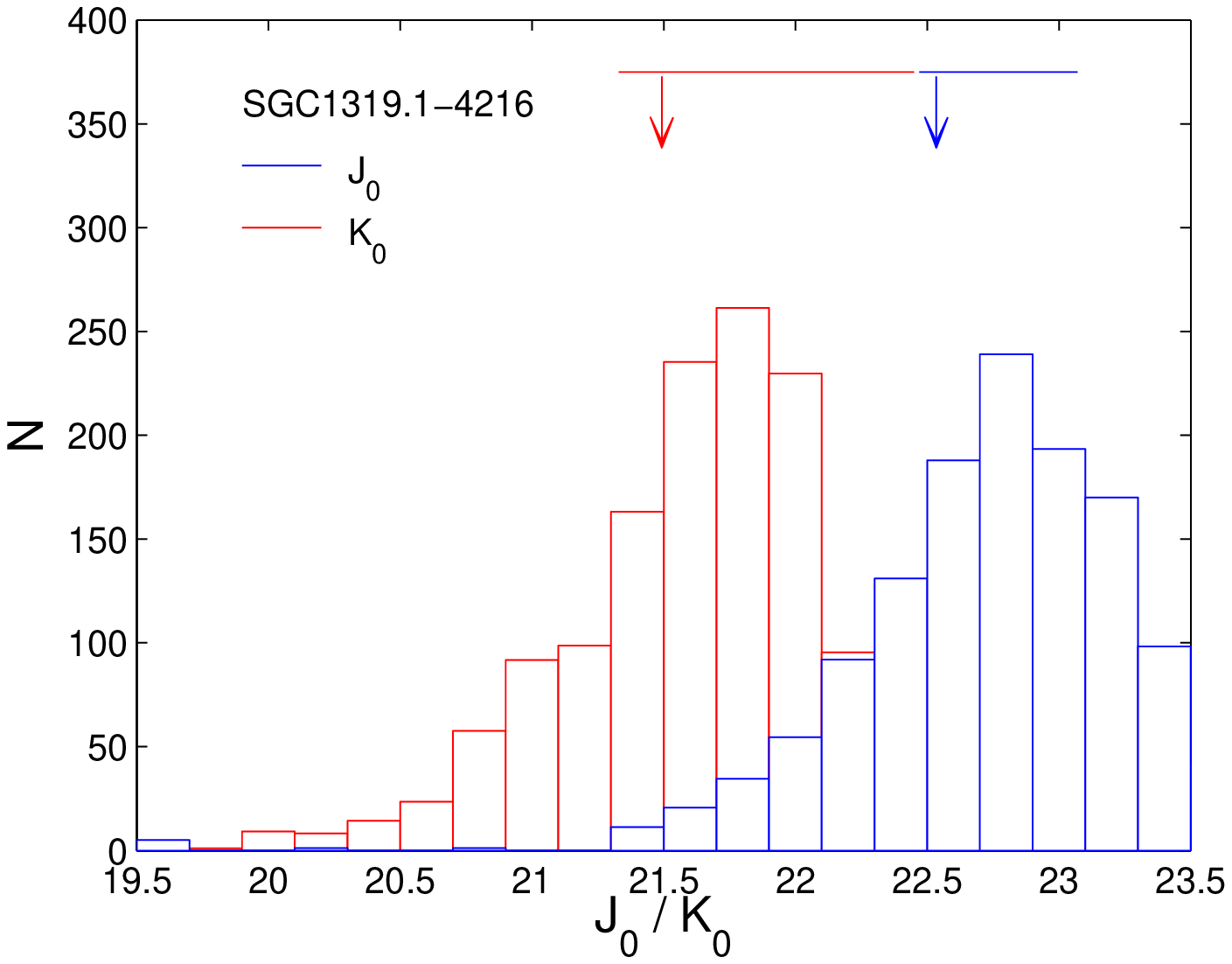}
  \includegraphics[width=7cm]{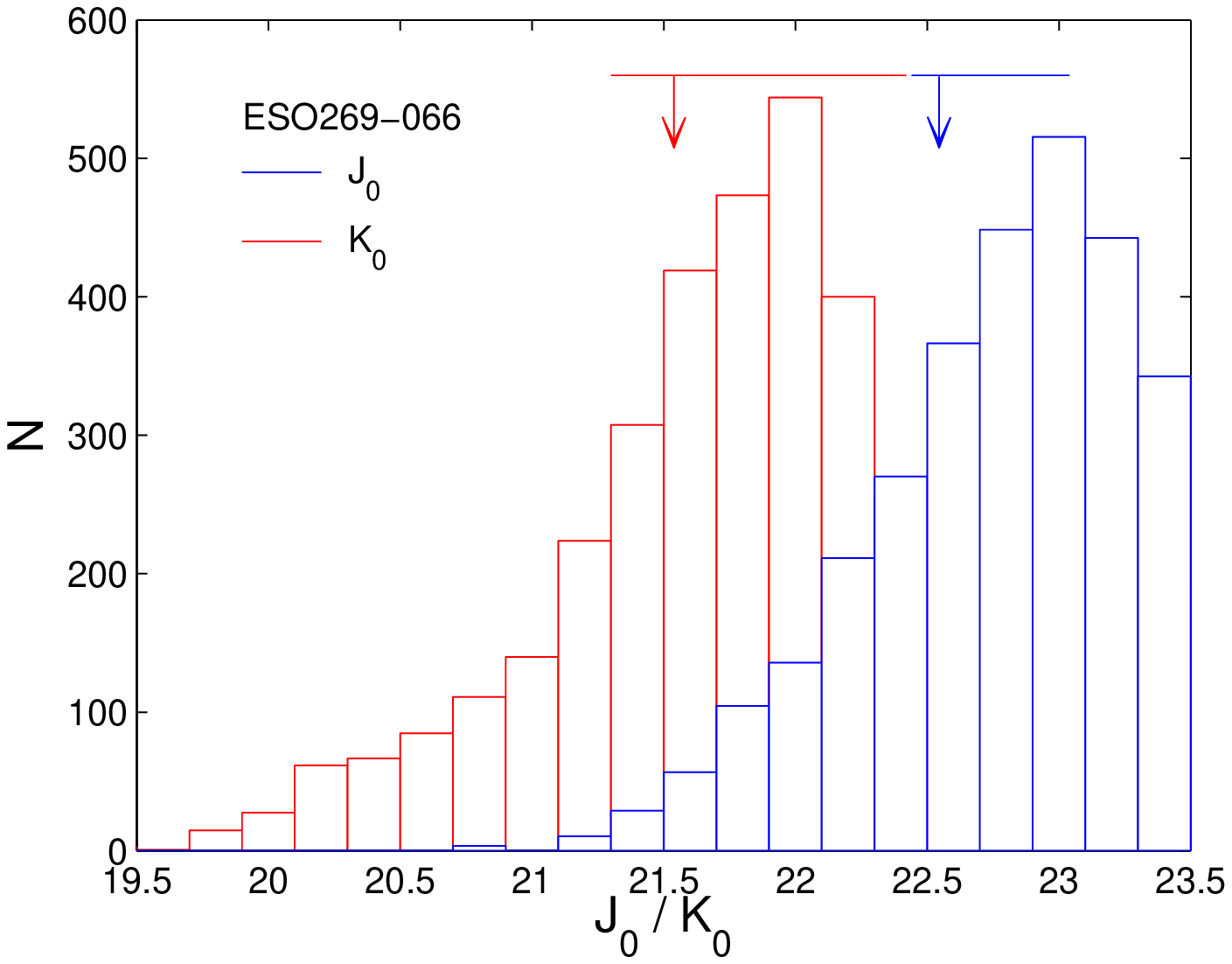}
 \caption{\footnotesize{NIR luminosity functions for the $J_0$- and $K_0$-bands of the target galaxies, ordered by increasing luminosity. The numbers have been corrected for Galactic foreground contamination. The arrows indicate the predicted values for the TRGB at the median metallicity of the galaxy, while the horizontal lines above the arrows show the range of predicted TRGB values that stem from the spreads in metallicity (see text for details).}}
 \label{lumfun}
\end{figure}

The NIR CMDs are presented in Fig. \ref{cmdnir}. We have corrected the magnitudes for foreground reddening, referring to the NED values derived from the Schlegel extinction maps \citep{schlegel98}. For all of the galaxies, the upper part of the RGB is visible at $J_0\gtrsim23$ and $K_0\gtrsim22$, and the stars above this limit are likely belonging to the IAPs of these galaxies. We also overplot the $50\%$ completeness limits in all of the panels.

We compute the expected TRGB magnitude in both bands using the formulae given in \citet{valenti04}, assuming the distance moduli derived in the previous Section and the median metallicities reported in Tab. \ref{infogen}. The resulting values are: $J_{0,TRGB}=22.96\pm0.18$, $22.56\pm0.18$ and $22.56\pm0.18$, and $K_{0,TRGB}=22.00\pm0.18$, $21.50\pm0.18$ and $21.52\pm0.18$ for CenA-dE1, SGC1319.1-4216 and ESO269-99, respectively. We emphasize that the TRGB is not constant as a function of metallicity in these bands, in other words its luminosity depends on the metallicity value of the galaxy and all of our targets have a considerable metallicity spread within them. However, as can be seen from the metallicity distribution functions presented in \citet{crnojevic10}, most of the stars in a galaxy have metallicity values around the median value, so that in the luminosity function of the galaxy the approximate locus of the TRGB will still be recognizable as a stellar count decrease toward brighter magnitudes (although not an as well-defined one, as would be the case in $I$-band). We thus check our NIR estimates of the approximate TRGB magnitudes by additionally plotting the luminosity function for both bands in Fig.~\ref{lumfun}. The luminosity functions were dereddened, and for each magnitude bin the number of predicted Galactic foreground stars from TRILEGAL (similar to that given by Besan\c{c}on) was subtracted. More precisely, we considered only a fraction of the predicted foreground stars in order to account for observational incompleteness effects (as a function of magnitude). As an example, for CenA-dE1 in the magnitude bin centered at $K_0=21$ the completeness (as derived from artificial star tests) is $\sim95\%$, so we subtract from the dwarf galaxy star counts $\sim95\%$ of the simulated TRILEGAL star counts in this magnitude bin. The expected TRGB values derived with the \citet{valenti04} formula using the median metallicities are shown in Fig. \ref{lumfun} as arrows, and agree well with the observations. We moreover show a horizontal line including the point of origin of the arrows in order to indicate the possible range of TRGB values stemming from the range of metallicities, again computed following the \citet{valenti04} TRGB calibration equation. I.e., if we use the lowest/highest end of a given galaxy's metallicity range to compute the TRGB with the \citet{valenti04} formula, we will find a correspondingly fainter/brighter TRGB value. In addition, the NIR CMDs suffer from much larger incompleteness and photometric errors than the optical CMDs, such that the most metal-poor tip of the RGB is fainter than the detection limit in CenA-dE1 and SGC 1319.1-4216, while it is close to the detection limit for ESO269-066. As a reference, in Fig. \ref{cmdnir} we also overplot stellar isochrones over the CMDs (as in Fig. \ref{cmdopt}) to indicate the metallicity range of the galaxies and the shape of the TRGB. We moreover draw a dashed line passing through the TRGB values computed from the lowest, the median and the highest metallicities found for each galaxy with the \citet{valenti04} formula, finding a good agreement with the theoretical isochrones.

Also the NIR CMDs are contaminated by Galactic foreground, but this time the luminous AGB region is not as heavily affected as it is in the optical observations (see right panel of Fig. \ref{cont}). The vertical feature extending from $J_0-K_0\sim0.3$ to $\sim1.0$ in Fig. \ref{cont} and \ref{cmdnir} and over the whole magnitude range is mainly due to Galactic old disk turnoff stars ($J_0-K_0\sim0.36$), Galactic RGB and red clump stars ($J_0-K_0\sim0.65$), and low-mass dwarfs with M$\leq0.6$M$_{\odot}$ ($J_0-K_0\sim0.85$, \citealt{girardi05}). The fact that these stars are distributed in vertical sequences is due to the range of distances and luminosities that they span. We will now take advantage of the fact that the Galactic foreground contamination is more easily recognizable in the NIR CMDs in order to look for AGB candidates in our galaxies.

\subsection{Combined CMDs} \label{cmdcomb_sec}

\begin{figure}
 \centering
  \includegraphics[width=9cm]{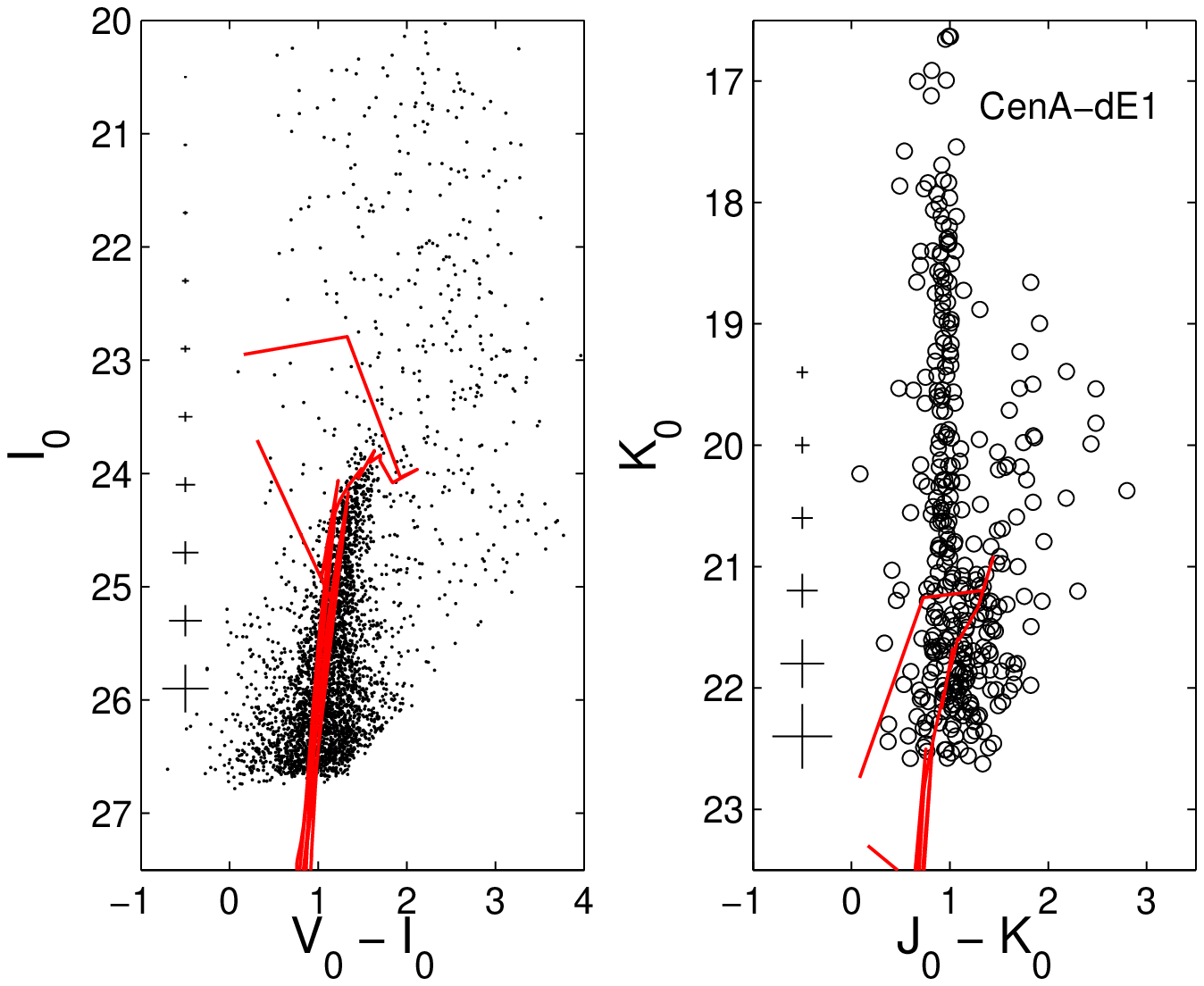}
  \includegraphics[width=9cm]{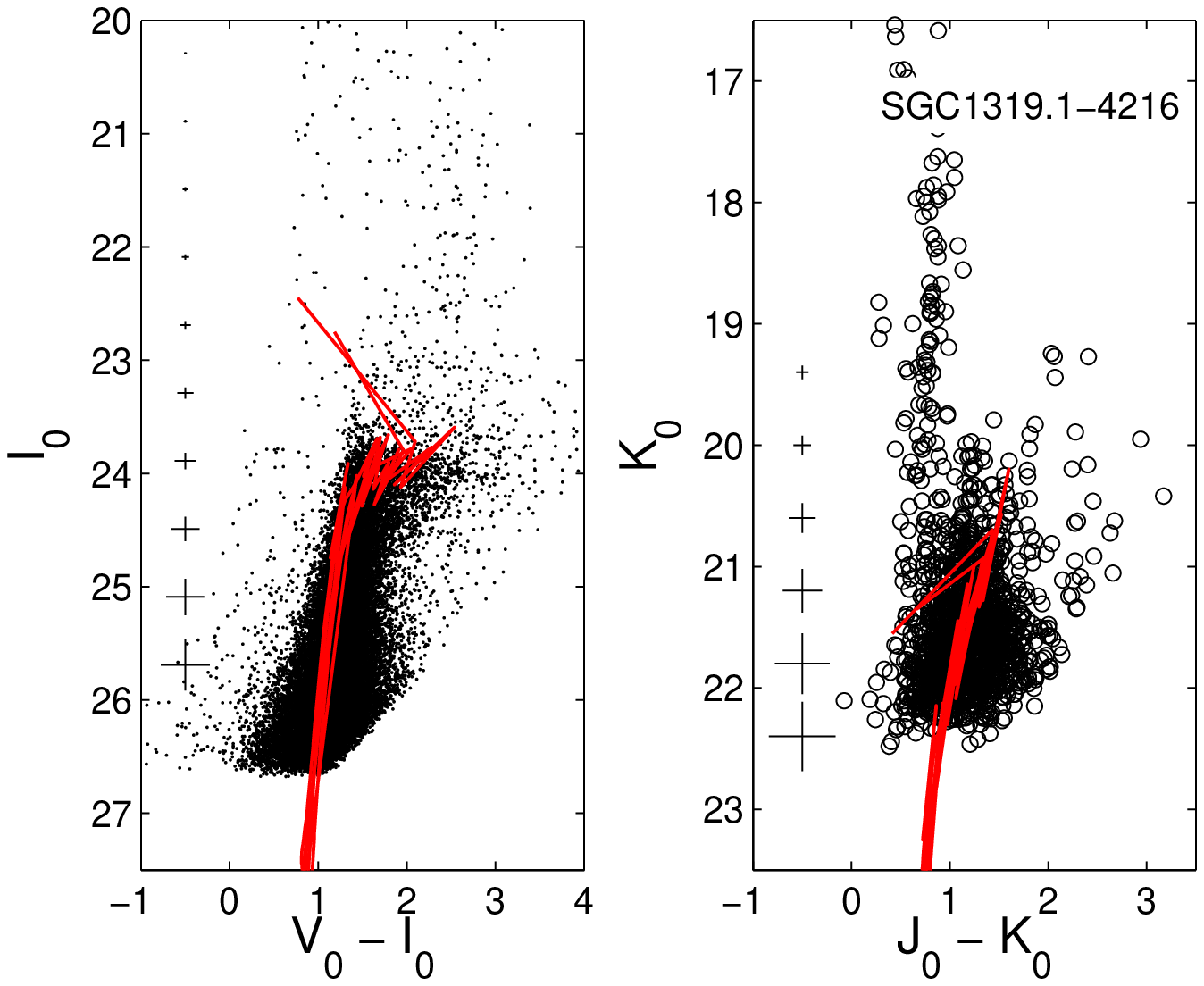}
 \caption{\footnotesize{Optical and NIR (dereddened) CMDs for CenA-dE1 and SGC1319.1-4216. Representative photometric errorbars are plotted along the CMDs. The red lines are isochrones from the Padova evolutionary models with improved AGB phases \citep{girardi10}. For CenA-dE1, they have the mean metallicity of the galaxy ([Fe/H]$\sim-1.5$) and ages of 2 and 9 Gyr; for SGC1319.1-4216 the metallicity is [Fe/H]$\sim-1.0$ and the ages 2 and 5 Gyr.}}
 \label{agbtracks}
\end{figure}

\begin{figure}
 \centering
  \includegraphics[width=9cm]{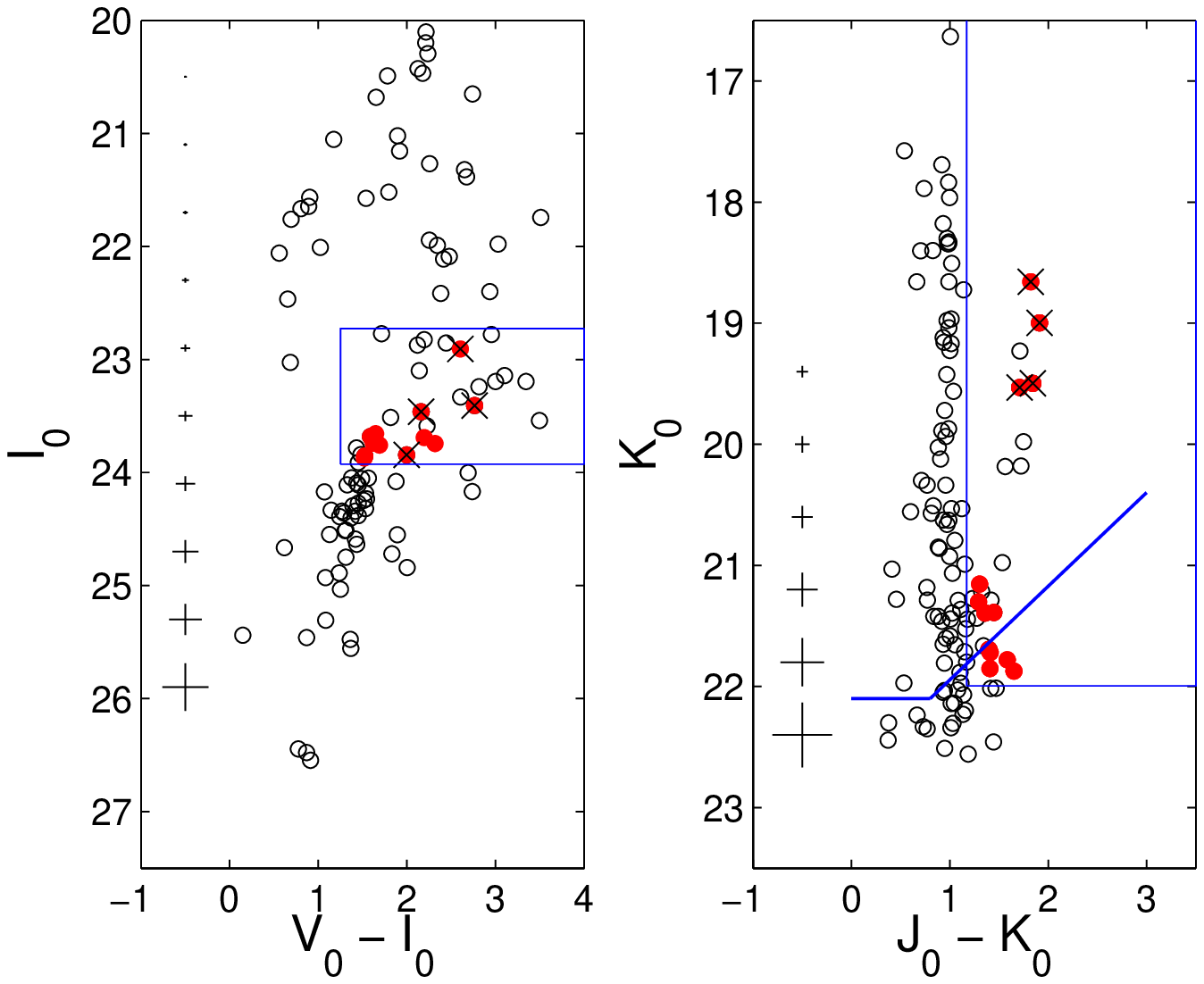}
  \includegraphics[width=8.5cm]{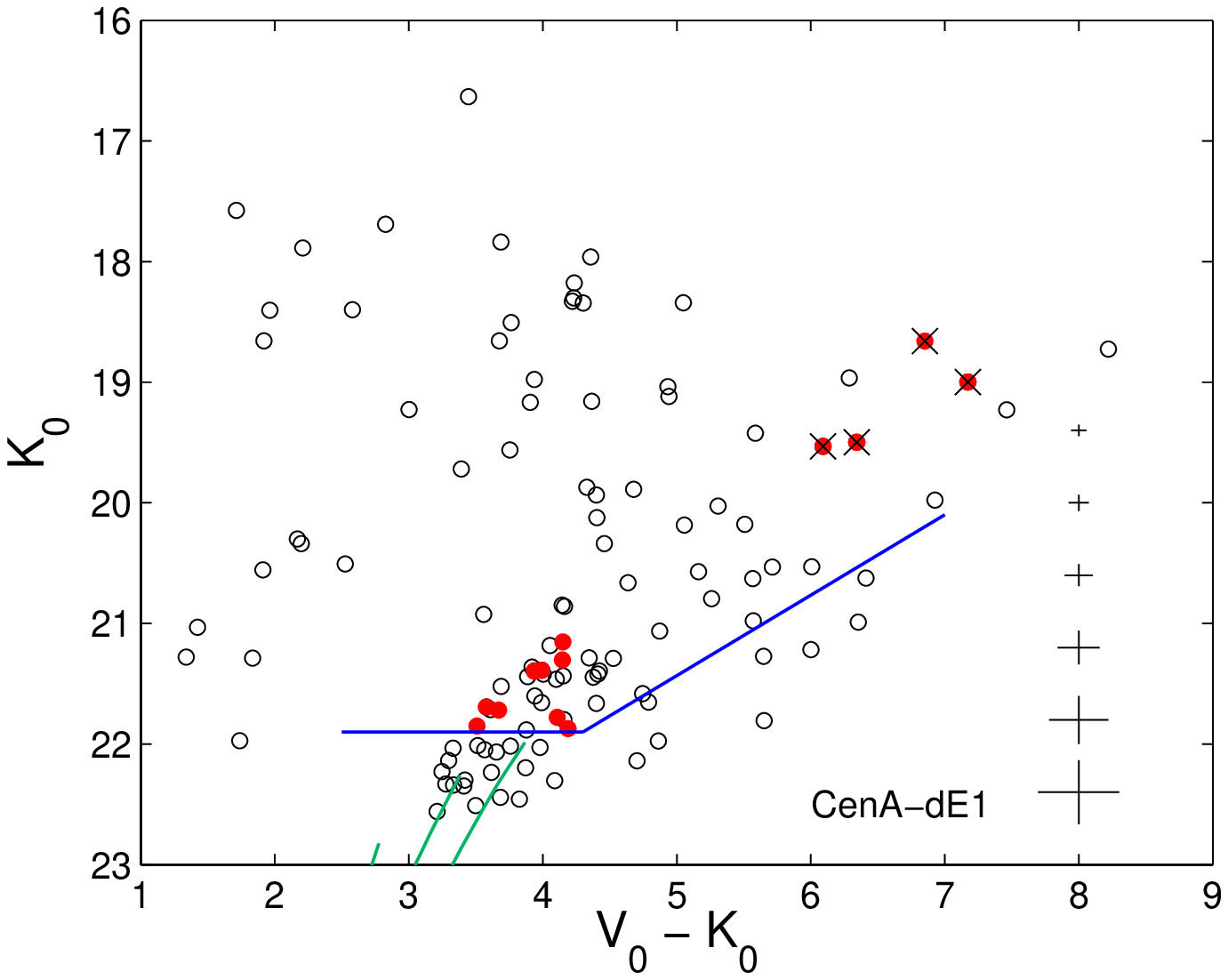}
 \caption{\footnotesize{\emph{Upper panel}. Optical and NIR (dereddened) CMDs for CenA-dE1, showing the stellar point sources cross-identified in both catalogs. Also shown are the boxes used to select luminous AGB candidate stars, and the resulting candidates are plotted as red circles. The objects marked with a black cross are cross-identified in the two datasets, but turn out to be probable background galaxies after visual inspection. Representative photometric errorbars are plotted along the CMDs. The blue solid lines in the NIR CMD indicate the $50\%$ completeness limit. \emph{Lower panel}. Combined CMDs in $V_0$- and $K_0$-bands for CenA-dE1. The stellar sources and the symbols are the same as above. We also plot (green lines) the isochrones with a fixed age of 10 Gyr and spanning the metallicity range of the galaxy, with the same values as in Fig. \ref{cmdopt}. The blue lines have the same meaning as in the upper panel.}}
 \label{cmdcomb1}
\end{figure}

\begin{figure}
 \centering
  \includegraphics[width=9cm]{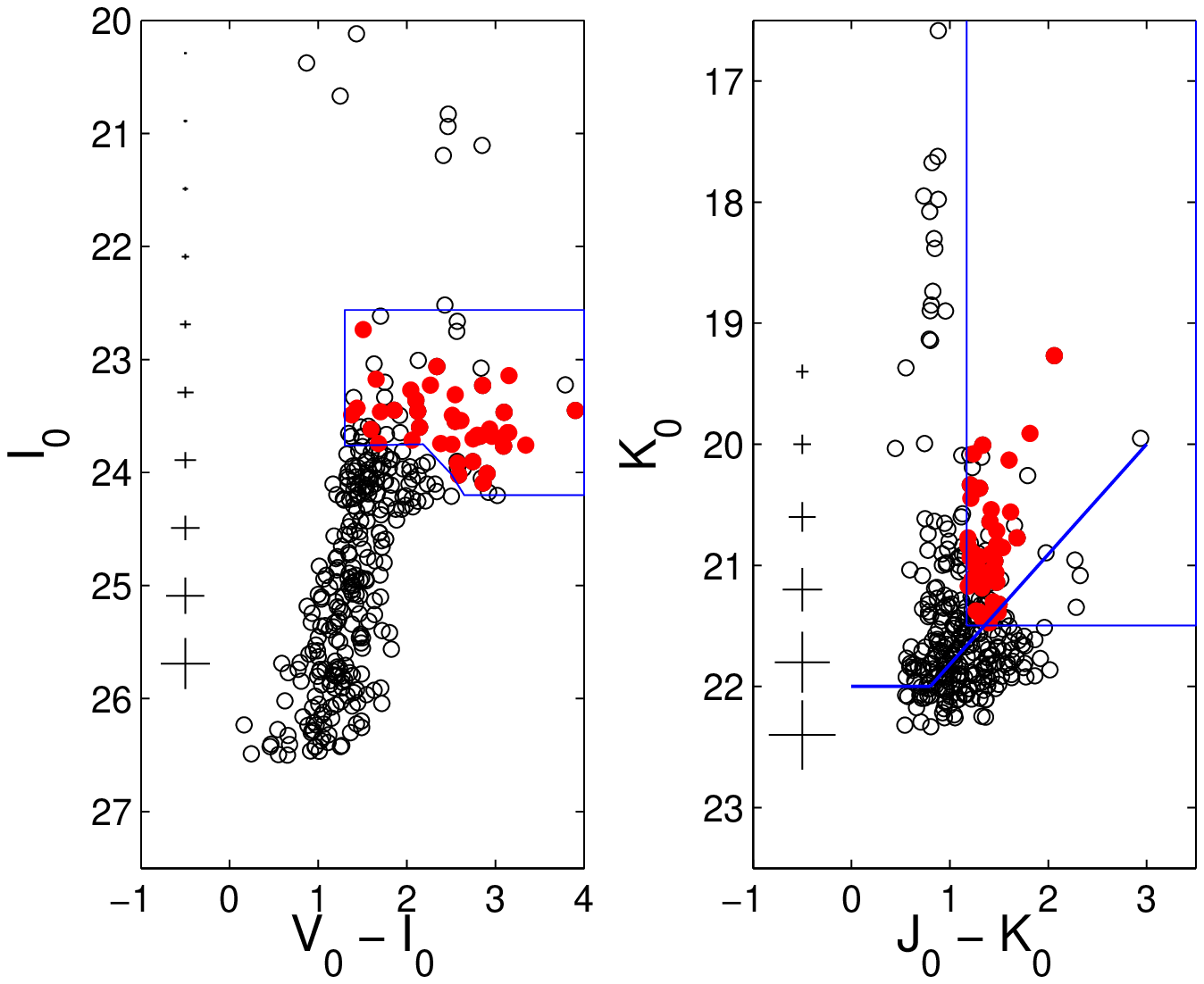}
  \includegraphics[width=8.5cm]{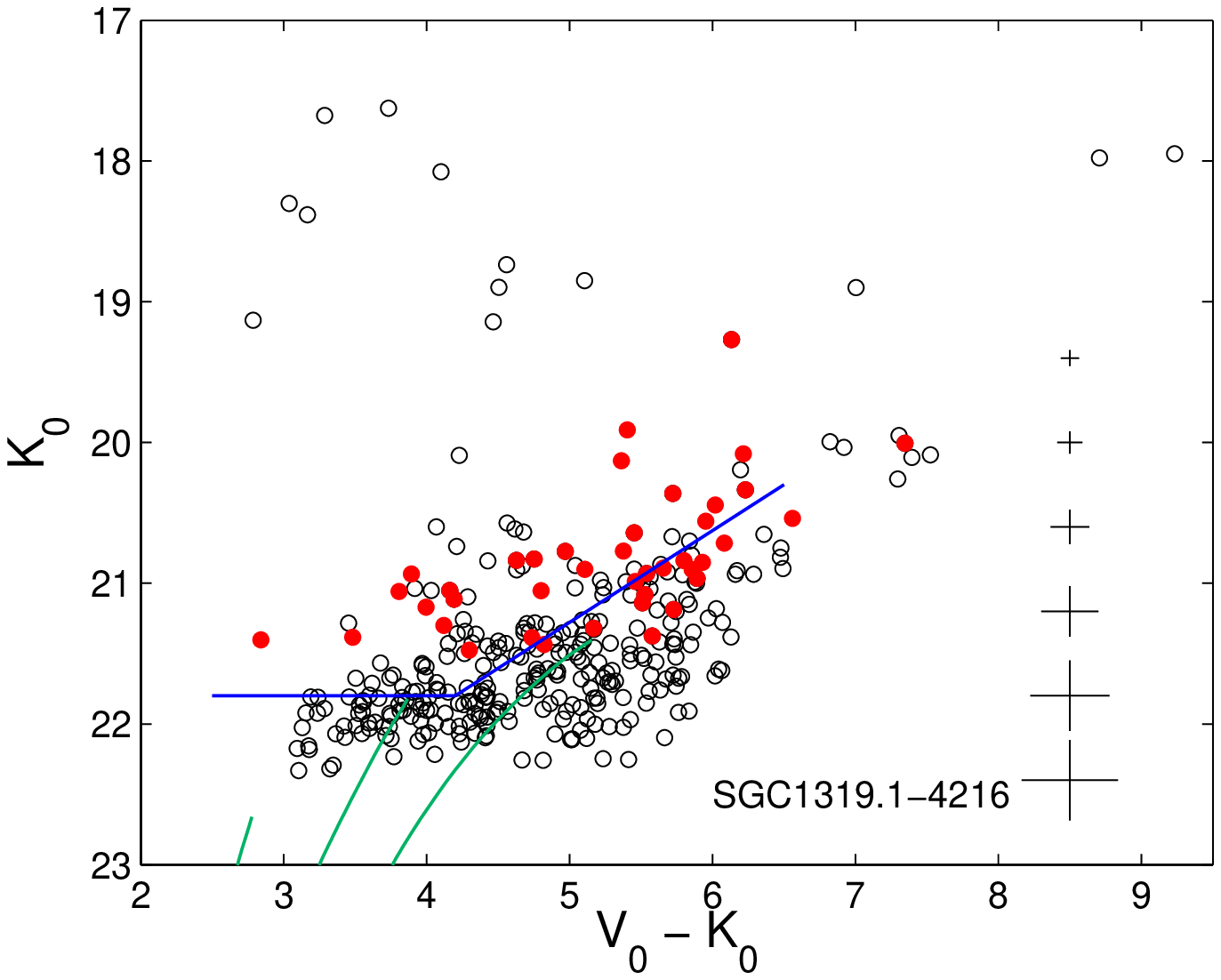}
 \caption{\footnotesize{Same as in Fig. \ref{cmdcomb1}, for the galaxy SGC1319.1-4216.}}
 \label{cmdcomb2}
\end{figure}

\begin{figure}
 \centering
  \includegraphics[width=9cm]{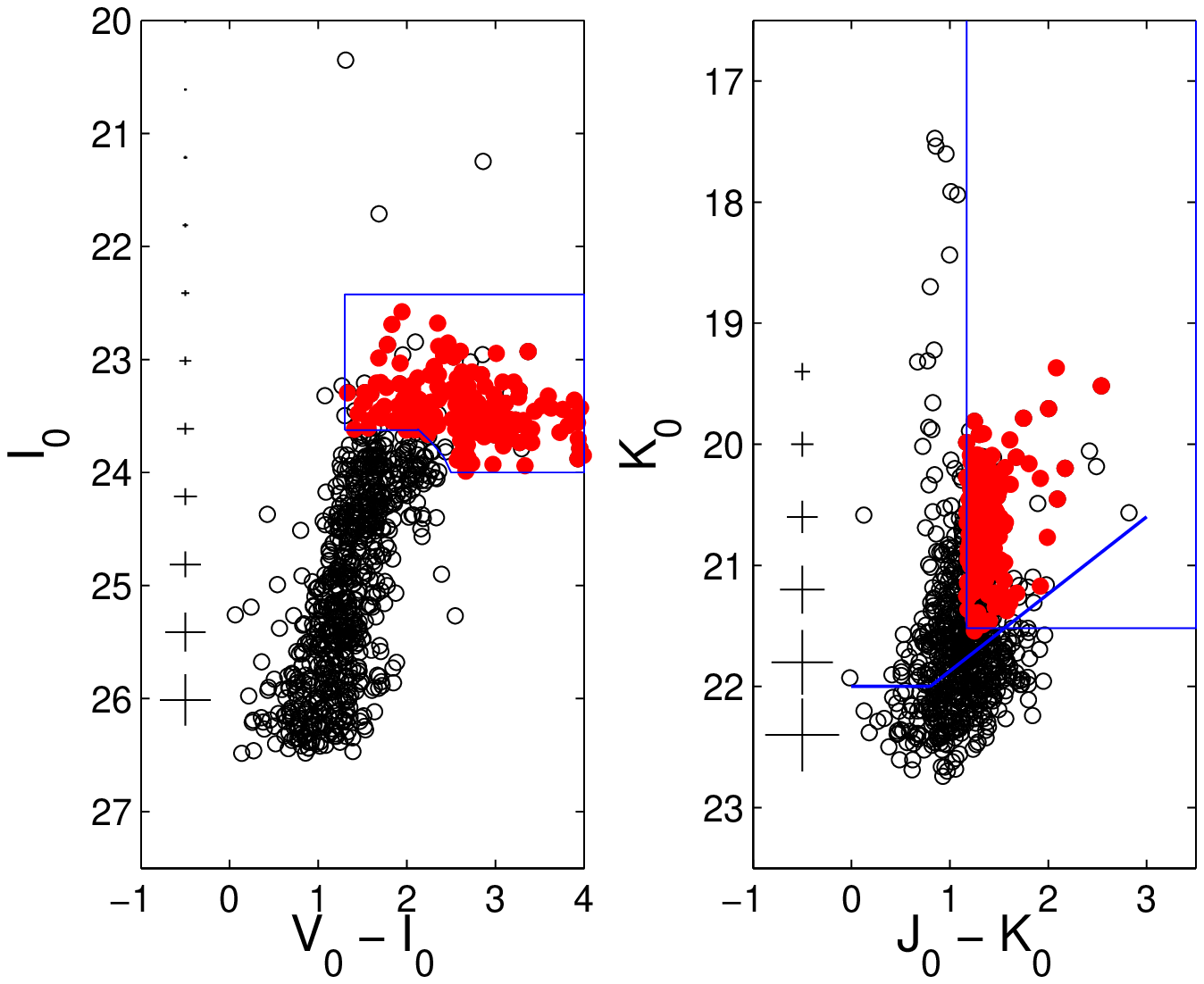}
  \includegraphics[width=8.5cm]{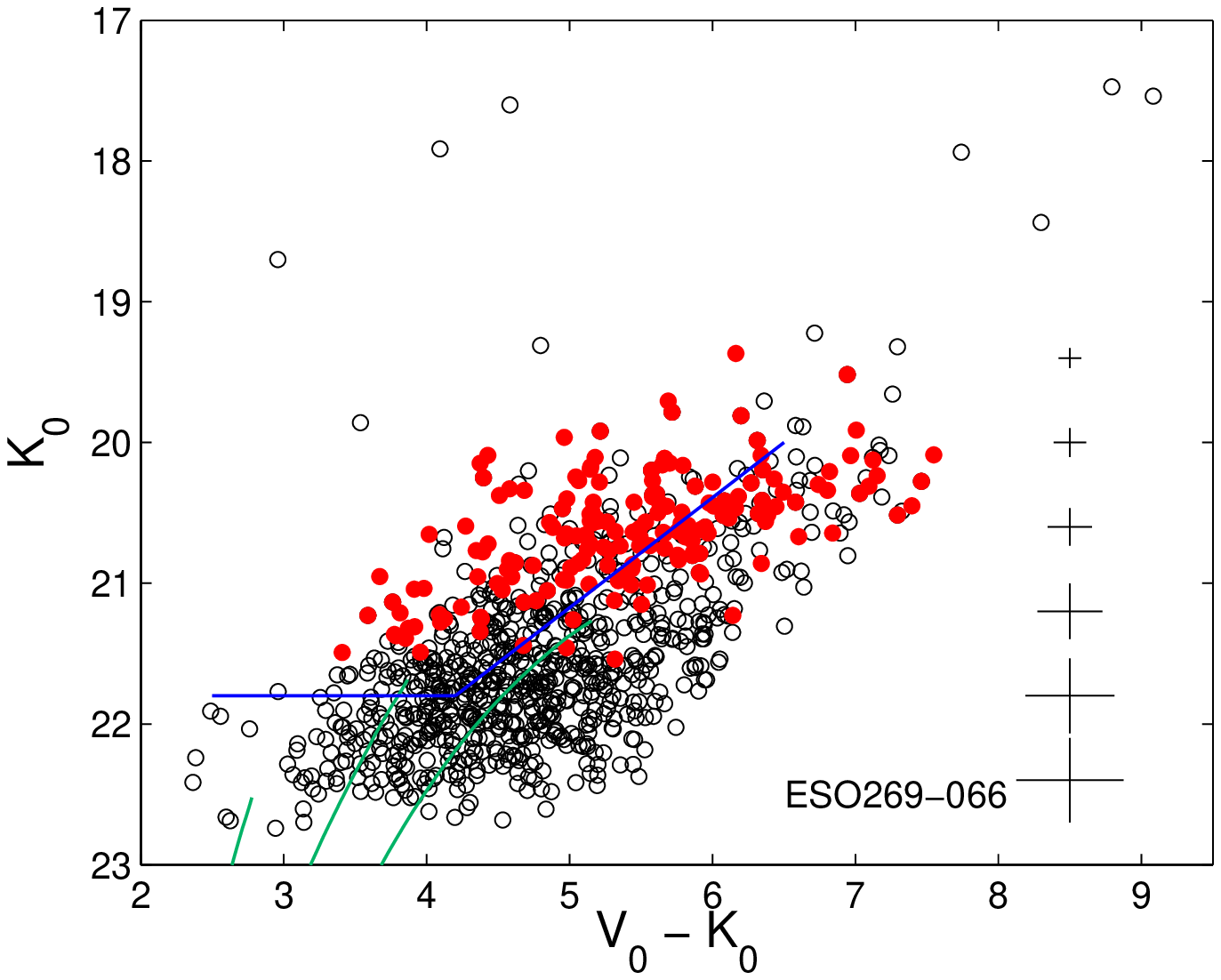}
 \caption{\footnotesize{Same as in Fig. \ref{cmdcomb1} and \ref{cmdcomb2}, for the galaxy ESO269-066.}}
 \label{cmdcomb3}
\end{figure}

Our goal is to use the combined HST/ACS and VLT/ISAAC data to reliably identify candidate AGB stars. To do this, we first cross-correlate the optical and NIR photometric catalogs. The coordinate transformations are initially derived using the IRAF tasks \emph{geomap} and \emph{geoxytran}. Successively, the combined catalog is obtained with the programme DAOMASTER \citep{stetson87}. In Fig. \ref{agbtracks} we show the expected location of luminous AGB stars, in both optical and NIR CMDs of CenA-dE1 and SGC1319.1-4216, by overplotting the isochrones from the Padova evolutionary models \citep[][]{girardi10}. We point out that some spread of the AGB star magnitudes and colors around the expected positions is due to photometric errors and to their intrinsic variability.

The combined lists of stars that have both optical and NIR photometry contain 115, 344, and 919 stars, for CenA-dE1, SGC1319.1-4216 and ESO269-066 respectively, and in the upper panels of Fig. \ref{cmdcomb1}, \ref{cmdcomb2} and \ref{cmdcomb3} we show both the optical and the NIR CMDs of the combined list of stars for each of the three galaxies. We note that these numbers are indeed much lower than the numbers of stars found in the optical photometric catalogs alone (see Sect. \ref{dataopt}). This is partly due to the fact that the ISAAC field of view is slightly smaller than the ACS one, and partly due to the higher resolution and photometric depth of ACS. Unfortunately, this results in a loss of information in the process of cross-correlation, since there could be some mismatches in the cases where more than one optical source is found in the vicinity of a NIR source. This is particularly true for SGC1319.1-4216 and ESO269-066, where the stellar density is higher, and from Fig. \ref{cmdcomb2} and \ref{cmdcomb3} we see that the NIR foreground sequences are poorly populated because of such mismatches. To avoid a loss of information for the part of the CMD in which we are most interested, we check visually the positions of the stars that are found within our selection boxes (see below) in both the NIR and optical catalogs, and add in this way a few more matches to our combined catalogs.

On the same CMDs, we also overplot the boxes used in both optical and NIR to select candidate AGB stars. For the optical, the box extends from a magnitude of $I_{0,TRGB}-\sigma_I$, where $\sigma_I$ is the observational error at that magnitude (in order not to select RGB stars with erroneously higher magnitudes because of photometric errors), up until $M_I\sim-5.5$. The color range goes from the bluest edge of the TRGB to $V_0-I_0\sim4$ (there are no stars redder than this value). This should be the magnitude range that luminous AGB stars with the average metallicity of the target galaxies cover \citep[see][ and references therein]{rejkuba06}. For SGC1319.1-4216 and ESO269-066, we choose to extend the optical selection box based on a few stars with colors $V_0-I_0\gtrsim2.5$ that have magnitudes below that of the TRGB derived from the median metallicity. These stars could belong to the metal-rich population, as the TRGB has in fact slightly fainter magnitudes at increasing metallicities (see, e.g., Fig. \ref{cmdopt}). For the NIR, the selection box contains all the stars that are found above the TRGB in these bands, and redwards of the reddest foreground stars of the vertical plume. Our final ``AGB candidates'' are those stars that are simultaneously found in the optical and the NIR selection boxes, and are plotted in red on these CMDs. We also overplot the $50\%$ completeness limits to the NIR CMDs, in order to point out that we might be missing the detection of a few luminous AGB stars with colors $J_0-K_0\gtrsim1.0-1.5$. This is true in particular for SGC1319.1-4216 and for ESO269-066, which have a more metal-rich extension than CenA-dE1. Given the number of stars in the low completeness region, we estimate that there could additionally be at least $\sim10$ undetected AGB stars for SGC1319.1-4216 and $\sim40$ for ESO269-066.

There are in total 13 AGB candidates for CenA-dE1, 41 for SGC1319.1-4216 and 176 for ESO269-066. For CenA-dE1, we can see that the four brightest candidates are detached from the rest of the red dots, so we perform a visual inspection of the images and conclude that they are barely resolved background galaxies. In general, the above reported numbers are lower limits to the total number of luminous AGB stars in the target galaxies, given the NIR incompleteness limits.

The lower panels of Fig. \ref{cmdcomb1}, \ref{cmdcomb2} and \ref{cmdcomb3} show the composite $V_0-K_0$ versus $K_0$ CMDs for the three studied galaxies. We plot the stars found in the combined lists and mark in red the AGB candidates. To show the expected TRGB position and to stress the metallicity spread of these galaxies, we also plot the same isochrones as in Fig. \ref{cmdopt} and \ref{cmdnir}. When looking at the combined CMDs, we notice that there are many objects that are found in the region above the TRGB, but that are not identified as candidate AGB stars. These sources are mostly foreground contaminants, which are distributed all over the CMDs for these magnitude combinations, but some of them might be unresolved background galaxies. In particular, \citet{saracco01} study deep ISAAC observations to derive the number counts of unresolved high-redshift galaxies. From their Fig. 1 we estimate a number of galaxies of up to $\sim300$ for our field of view, for magnitudes $18<K_0<22$ and distributed on the CMD as shown in their Fig. 3. Due to the combination of high resolution HST data in addition to excellent seeing ISAAC images, as well as to the quality cuts applied to our PSF photometry, most of these galaxies will be rejected from the final combined catalogs, but there is still the possibility of having a few high-redshift and compact contaminants among our AGB candidates, although this cannot be clarified with the current data. The objects found at colors $V_0-K_0\gtrsim8$ are indeed barely resolved background galaxies, as confirmed by visual inspection.

\begin{table*}
 \centering
\caption{List of the AGB candidates for each galaxy in this study.}
\label{agb_list}
\begin{tabular}{lccccccc}
\hline
\hline
Galaxy&ID&$\alpha_{2000}$&$\delta_{2000}$&$J_0$&$K_0$&$V_0$&$I_0$\\
\hline
\object{CenA-dE1}&$168$&$13\;12\;46.07$&$-41\;50\;53.23$&$22.60\pm0.10$&$21.30\pm0.07$&$25.45\pm0.03$&$23.76\pm0.03$\\
&$204$&$13\;12\;49.83$&$-41\;50\;38.48$&$23.13\pm0.14$&$21.72\pm0.08$&$25.39\pm0.03$&$23.87\pm0.03$\\
&$329$&$13\;12\;47.39$&$-41\;50\;35.96$&$22.75\pm0.10$&$21.39\pm0.07$&$25.33\pm0.03$&$23.71\pm0.03$\\
&...&&&&&&\\
\object{SGC1319.1-4216}&$224$&$13\;22\;01.21$&$-42\;32\;31.06$&$22.77\pm0.10$&$21.40\pm0.12$&$24.24\pm0.02$&$22.74\pm0.02$\\
&$517$&$13\;22\;01.97$&$-42\;32\;15.43$&$22.88\pm0.13$&$21.48\pm0.12$&$25.77\pm0.05$&$23.71\pm0.03$\\
&$562$&$13\;22\;03.12$&$-42\;32\;17.90$&$21.66\pm0.05$&$20.45\pm0.05$&$26.46\pm0.06$&$23.67\pm0.03$\\
&...&&&&&&\\
\object{ESO269-066}&$919$&$13\;13\;10.34$&$-44\;52\;49.72$&$21.82\pm0.05$&$20.51\pm0.03$&$25.65\pm0.04$&$23.47\pm0.02$\\
&$1008^*$&$13\;13\;11.16$&$-44\;53\;30.06$&$22.23\pm0.07$&$20.67\pm0.05$&$26.17\pm0.05$&$23.61\pm0.03$\\
&$1246$&$13\;13\;09.19$&$-44\;53\;01.55$&$22.13\pm0.07$&$20.73\pm0.05$&$26.29\pm0.06$&$23.46\pm0.02$\\
&...&&&&&&\\
\hline
\end{tabular}
\tablefoot{The columns contain: (1): galaxy name; (2): stellar ID from the combined catalog; (3-4): equatorial coordinates (J2000, units of right ascension are hours, minutes, and seconds, and units of declination are degrees, arcminutes, and arcseconds); (5) dereddened apparent $J_0$-band magnitude; (6): dereddened apparent $K_0$-band magnitude; (7): dereddened apparent $V_0$-band magnitude; (7): dereddened apparent $I_0$-band magnitude. Note that the given photometric errors are the ones resulting from the photometric package. Variable candidates are denoted with an asterisk. For each galaxy, only the first three entries are reported, while the complete table is available from the CDS.}
\end{table*}

%________________________________________________________________

\section{Intermediate-age populations} \label{agb_sec}

\begin{figure}
 \centering
  \includegraphics[width=8cm]{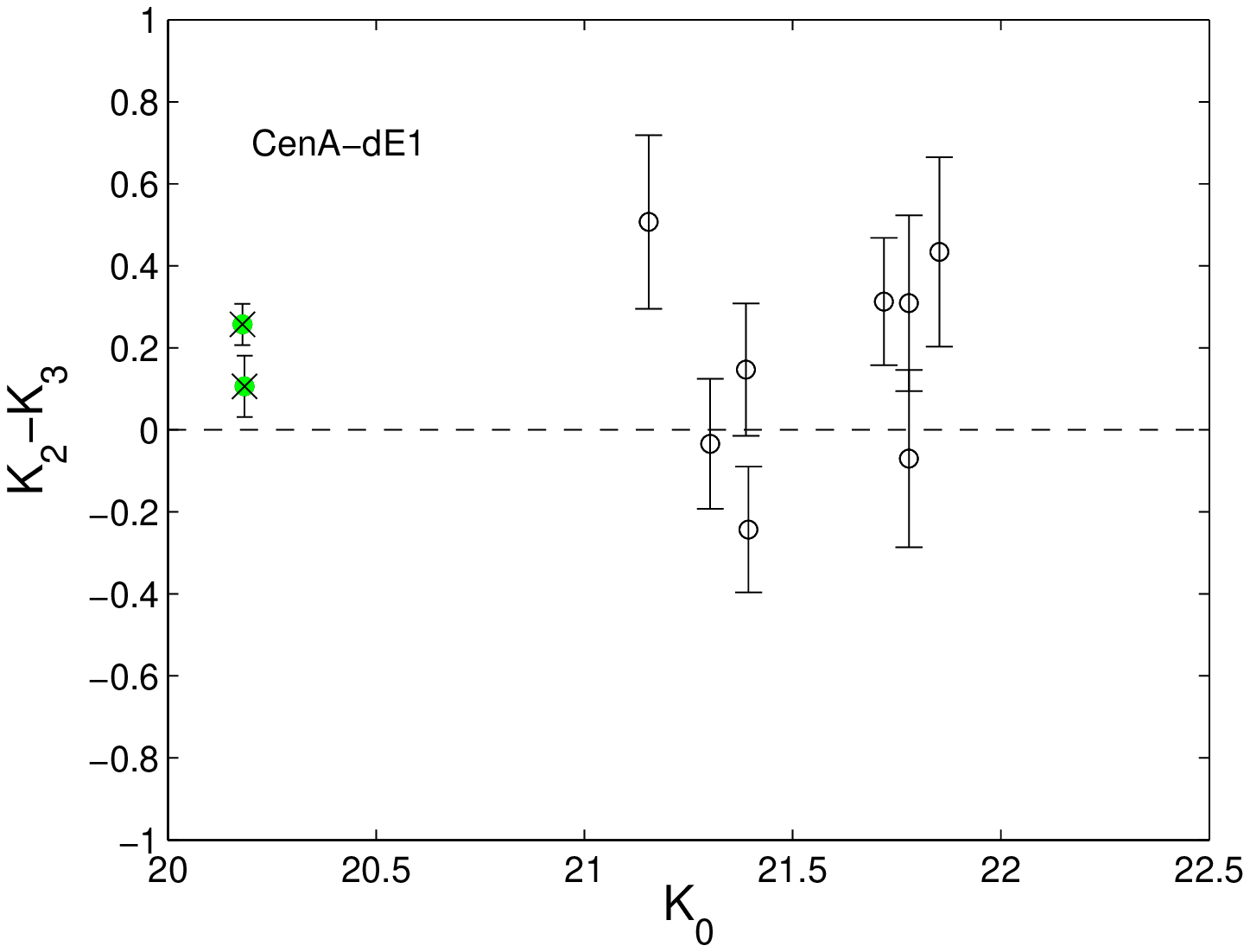}
  \includegraphics[width=8cm]{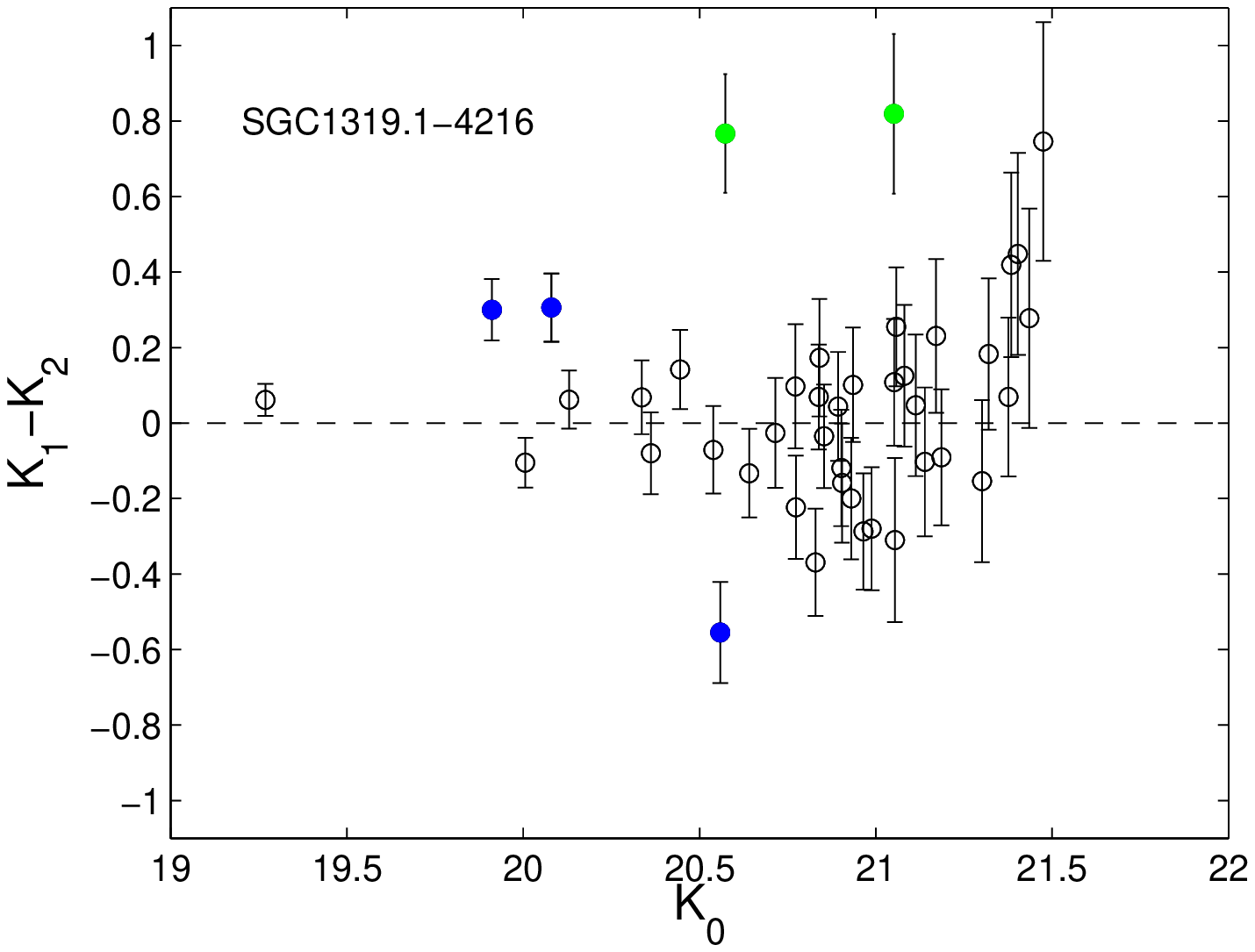}
  \includegraphics[width=8cm]{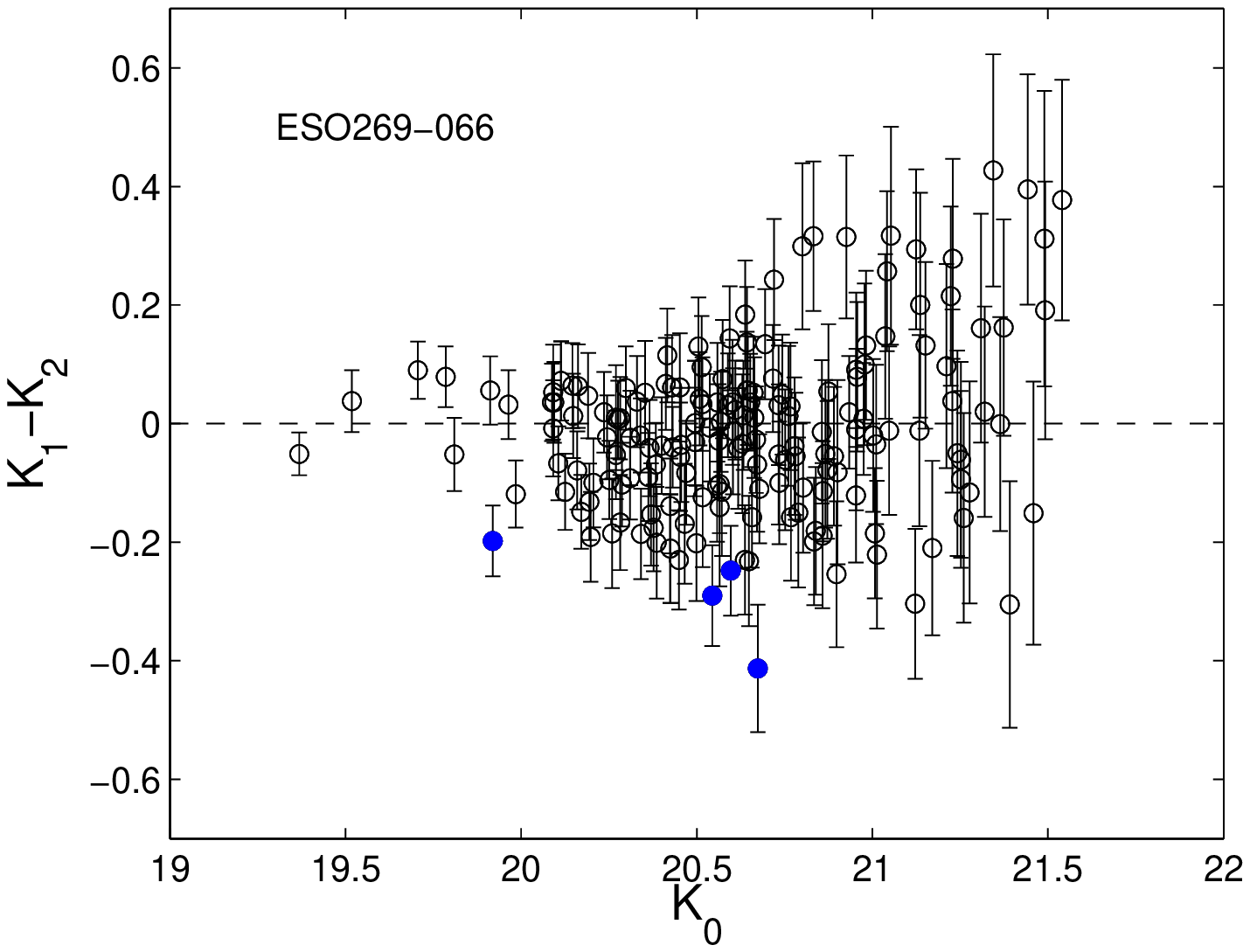}
 \caption{\footnotesize{$K$-band magnitude difference between the first and second observation (or second and third in the case of CenA-dE1), against the combined dereddened $K$ magnitude. The target galaxies are ordered by increasing luminosity. We display the magnitude difference for candidate AGB stars, plus for those stars that lie at the edges of the selection boxes but show variability (green dots). For CenA-dE1, the latter turn out to be probable background galaxies after visual inspection and have a black cross on top of the circle (see text for details). Blue objects are AGB candidates for which the magnitude varies by more than 3 times the combined photometric errors of the individual measurements in the two observations (or in any combination of observations for CenA-dE1). Finally, both for CenA-dE1 and SGC1319.1-4216 one AGB candidate is not shown since it has a bad measurement in one of the two plotted $K$-bands.\vspace{1cm}}}
 \label{variab}
\end{figure}

From the combined optical and NIR CMD analysis of the previous Section, we are left with 9 luminous AGB candidates for CenA-dE1, 41 for SGC1319.1-4216 and 176 for ESO269-066. We report the coordinates and magnitudes of these candidate AGB stars in Tab. \ref{agb_list} (the complete table is available electronically). These stars belong to the IAP of the target dwarfs, while a few old and metal-rich AGB stars could be also present for SGC1319.1-4216 and ESO269-066, given the red extension of their RGBs. We stress that the old and metal-rich stars in our sample are a very small percentage of the whole luminous AGB population, given the low fraction of stars with [Fe/H]$\lesssim-0.7$. Therefore, in the subsequent analysis, we will neglect their contribution, thus considering all our AGB candidates to belong to the IAP. We now want to look at their properties in more detail.

Generally speaking, with these data it is not possible to firmly separate carbon-rich from oxygen-rich stars among our AGB candidates. However, for such metal-poor galaxies we would expect to find carbon-rich stars at colors $J_0-K_0\gtrsim1.5$ \citep[e.g.][and references therein]{kang06}, and a few stars with these colors are indeed present in all of our target galaxies. 

We check whether our stellar samples contain dust enshrouded AGB stars. This kind of objects are extremely faint or undetected in the optical, very red at NIR wavelengths and thus not easily detectable in the $J$-band because of incompleteness effects in our observations. For example, \citet{vanloon05} consider a sample of $\sim40$ stellar clusters with a range of ages and metallicities in the Small and Large Magellanic Clouds. They find a total of about 30 dust enshrouded AGB stars in $\sim20$ young and intermediate-age clusters. These stars are found at $J-K>2.5$, and have metallicities higher than [Fe/H]$=-0.9$ dex. However, for clusters with ages and metallicities comparable to our target galaxies, no dust enshrouded AGB stars were detected. We thus do not expect a significant number of dust enshrouded stars to be present in our target galaxies. We search for stars that have good $K$-band measurement but no $J$-band counterpart, and find two such objects in CenA-dE1, none in SGC1319.1-4216 and one in ESO269-066. Of the mentioned sources, in CenA-dE1 one is found slightly outside the limiting radius, while the second is close to the center but has a good measurement only for one of the two $K$-bands; in ESO269-066 the dust enshrouded candidate also has a bad measurement in one of the two bands. We thus mention that these are candidates but could just as well be unresolved background galaxies (see previous Sect.).

\subsection{Variable candidates} \label{var_sec}

We can also look for additional AGB candidates by considering variability, which is an intrinsic characteristic of luminous AGB stars. For all of the target galaxies we have at least two observations in the $K$-band, so we use the difference between the stellar magnitudes at different epochs as a variability indicator. For a long period variable star, the typical maximum magnitude difference is $\sim0.1$ to $\sim1.5$ mag in the $K$-band, and the period is on the order of $\sim10^{2-3}$ days \citep[see for example][and references therein]{rejkuba03}. We should thus expect to see variations of a few tens of a magnitude at most, given the observing timescales for our targets (see Tab. \ref{infonir}).

For CenA-dE1, there are three observations in the $K$-band due to one repeated observation. There are 36 days between the first and the last one (see Tab. \ref{infonir}), which are barely enough to put a lower limit on the number of possible long-period variables. We check whether there are variations between the different $K$-band observations, but find none. We then also check the whole combined list of sources, looking for stars that display a magnitude variation of more than 3 times the combined photometric errors of the individual measurements. We find two additional variable sources that lie just below the lower limits of the AGB selection boxes, and thus include them in our AGB candidates list. However, when checking them on the images we find that their profiles look like those of barely resolved background galaxies. In Fig. \ref{variab} (upper panel) we display the $K$-band magnitude difference between the second and the third epochs, since these are the ones with better seeing, for all the candidate AGB stars except one, because it has a bad measurement in the second $K$-band observation. Shown (in green) are also the two likely background galaxies.

The $K$-band observations of SGC1319.1-4216 were taken 57 days apart, and three of the AGB candidates display variability (blue dots in the central panel of Fig. \ref{variab}). When considering the entire sample, two stars that lie just leftwards of the NIR selection box, and are found inside the optical selection box, are indeed variables exhibiting a luminosity change by more than $3\sigma$ (green symbols). We add the two latter to the number of candidate AGB stars for SGC1319.1-4216 (and report them in the electronic version of Tab. \ref{agb_list}). Also in this case, one AGB candidate has a bad measurement in the second $K$-band observation and is thus not shown in the plot.

Finally, for ESO269-066 the timescale for the observations is very short, only 20 days. We are able to see a variability of more than $3\sigma$ for only four AGB candidates, as can be seen from the bottom panel of Fig. \ref{variab}. No additional variable candidates were found for this galaxy.

\subsection{Absolute bolometric magnitudes} \label{age_sec}

We can now use the luminosity of the candidate AGB stars to constrain their ages. This is possible since in this evolutionary stage, at a given metallicity, there is no degeneracy between position in the CMD and age. In particular, the maximum bolometric luminosity in the sample of AGB candidates can tell us when the most recent episode of star formation took place in a galaxy, with the further advantage that the metallicity dependence of the luminosity is weak compared to the age dependence. We refer in particular to Fig. 19 of \citet{rejkuba06}, where an empirical relationship is found between age and absolute bolometric magnitude of the tip of the AGB, starting from data for LMC and SMC clusters. Although the bolometric luminosity at the AGB tip also depends on metallicity, this has less of an effect than the age dependence. For example, according to the Padova stellar evolutionary models, at a fixed age (either 2, 4, 6 or 9 Gyr) the difference in bolometric AGB tip luminosity between metallicities of Z=0.0006 (i.e., [Fe/H]$=-1.5$), Z=0.0019 (i.e., [Fe/H]$=-1.0$) and Z=0.006 (i.e., [Fe/H]$=-0.5$) is only $\sim0.2$ mag.

We thus first compute the bolometric magnitudes of our luminous AGB candidates for each galaxy. We can apply bolometric corrections to both our optical and NIR results, and see whether they give consistent values. Following \citet{rejkuba06} (see also their discussion), we adopt NIR bolometric corrections from \citet{costa96} and optical ones from \citet{dacosta90}. The results for the three most luminous stars in both optical and NIR wavelengths (which in some cases overlap) are reported in Tab. \ref{bol}.

For CenA-dE1, the three most luminous stars in the NIR have absolute bolometric magnitudes of $M_{bol,NIR}\sim-4.1\pm0.2$ to $\sim-3.8\pm0.2$, while the values derived from the optical are $M_{bol,opt}\sim-4.1\pm0.2$ to $\sim-4.0\pm0.2$ (with one star in common between the two subsamples). The optical data were taken about 20 days later than the NIR data. We take the average $M_{bol}$ from the three most luminous AGB candidates, which are $M_{bol,NIR}\sim-3.9\pm0.4$ and $M_{bol,opt}\sim-4.1\pm0.3$, respectively. When putting these values on the age-absolute bolometric magnitude of AGB tip relation of \citet{rejkuba06}, we obtain an age of approximately $9.0\pm1$ Gyr for the most recent episode of star formation, indicating that there was very little activity in this galaxy other than at quite old ages.

\begin{table*}
 \centering
\caption{Magnitudes of the most luminous (in bolometric magnitude) AGB candidates for each galaxy.}
\label{bol}
\begin{tabular}{lccccccc}
\hline
\hline
Galaxy&ID&$I_0$&$K_0$&$V_0-I_0$&$J_0-K_0$&$M_{bol,opt}$&$M_{bol,NIR}$\\
\hline
\object{CenA-dE1}&$420$\tablefootmark{a}&$23.69\pm0.03$&$21.78\pm0.11$&$2.20$&$1.58$&$-4.14\pm0.18$&$-3.24\pm0.24$\\
&$9323$\tablefootmark{a}&$23.74\pm0.03$&$21.87\pm0.20$&$2.32$&$1.65$&$-4.11\pm0.18$&$-3.10\pm0.25$\\
&$9248$\tablefootmark{a,b}&$23.66\pm0.02$&$21.15\pm0.08$&$1.65$&$1.30$&$-4.04\pm0.18$&$-4.08\pm0.22$\\
&$168$\tablefootmark{b}&$23.76\pm0.02$&$21.30\pm0.07$&$1.69$&$1.29$&$-3.95\pm0.18$&$-3.94\pm0.21$\\
&$329$\tablefootmark{b}&$23.71\pm0.03$&$21.39\pm0.07$&$1.62$&$1.36$&$-3.98\pm0.18$&$-3.79\pm0.22$\\
\object{SGC1319.1-4216}&$224$\tablefootmark{a}&$22.74\pm0.02$&$21.40\pm0.12$&$1.51$&$1.36$&$-4.76\pm0.17$&$-3.61\pm0.28$\\
&$1361$\tablefootmark{a,b}&$23.14\pm0.02$&$20.08\pm0.04$&$3.15$&$1.24$&$-4.75\pm0.17$&$-5.05\pm0.27$\\
&$1316$\tablefootmark{a,b}&$23.06\pm0.02$&$19.27\pm0.02$&$2.34$&$2.06$&$-4.64\pm0.17$&$-5.37\pm0.48$\\
&$1482$\tablefootmark{b}&$23.45\pm0.02$&$20.00\pm0.03$&$3.90$&$1.34$&$-4.63\pm0.17$&$-5.03\pm0.28$\\
\object{ESO269-066}&$3689$\tablefootmark{a}&$22.58\pm0.01$&$20.15\pm0.04$&$1.94$&$1.22$&$-4.89\pm0.17$&$-4.86\pm0.32$\\
&$2139$\tablefootmark{a}&$22.68\pm0.01$&$20.24\pm0.04$&$2.35$&$1.31$&$-4.89\pm0.17$&$-4.58\pm0.34$\\
&$4406$\tablefootmark{a,b}&$22.93\pm0.02$&$19.99\pm0.03$&$3.37$&$1.17$&$-4.88\pm0.17$&$-5.07\pm0.32$\\
&$3774$\tablefootmark{b}&$22.93\pm0.02$&$19.37\pm0.18$&$2.60$&$2.08$&$-4.70\pm0.17$&$-5.13\pm0.63$\\
&$3170$\tablefootmark{b}&$23.21\pm0.02$&$19.92\pm0.03$&$1.92$&$1.30$&$-4.25\pm0.17$&$-5.01\pm0.34$\\
\hline
\end{tabular}
\tablefoot{
The columns contain: (1): galaxy name; (2): stellar ID from the combined catalog; (3): dereddened apparent $I_0$-band magnitude; (4): dereddened apparent $K_0$-band magnitude; (5): dereddened optical color; (6): dereddened NIR color; (7) absolute bolometric magnitude computed from optical values; and (8) absolute bolometric magnitude computed from NIR values. Note that the combined (ISAAC and ACS) and dereddened NIR and optical magnitudes are given with the photometric errors resulting from the photometric package.\\
\tablefoottext{a}{Among the three most luminous objects in the optical.}\\
\tablefoottext{b}{Among the three most luminous objects in the NIR.}
}
\end{table*}

In SGC1319.1-4216, the three most luminous AGB candidates have $M_{bol,NIR}\sim-5.4\pm0.5$ to $\sim-5.0\pm0.3$ and $M_{bol,opt}\sim-4.8\pm0.2$ to $\sim-4.6\pm0.2$. Two out of three stars are the same for the two samples, and the optical data were taken between the two $K$-band observations. The average values are $M_{bol,NIR}\sim-5.1\pm0.6$ and $M_{bol,opt}\sim-4.7\pm0.3$, respectively. These bolometric magnitudes lead to ages of $\sim2.0\pm1.5$ Gyr and $\sim5.0\pm2.0$ for the last star formation episode.

Finally, for ESO269-066 the three highest luminosity values found in the NIR range from $M_{bol,NIR}\sim-5.1\pm0.6$ to $\sim-5.0\pm0.3$, while in the optical they all are $M_{bol,opt}\sim-4.9\pm0.2$ (with one common star out of the three most luminous between the subsamples). This galaxy was observed with HST three months later than in the NIR bands. As before we compute the average values for the NIR and optical, $M_{bol,NIR}\sim-5.1\pm0.8$ and $M_{bol,opt}\sim-4.9\pm0.3$, and derive from these values the time of the most recent star formation, resulting in $2.0\pm1.5$ Gyr and $3.0\pm1.5$ Gyr ago.

We just stress that we may expect the bolometric values to differ somewhat between the optical and NIR results, given that the bolometric corrections may have differences from one set of bands to the other, and also considering the fact that our sources are possibly long-period variable stars. However, in all of the cases we find a very good agreement between the two datasets. We also point out that the optical bolometric corrections from \citet{dacosta90} are based on stars with colors $V-I<2.6$, so for the three objects in SGC1319.1-4216 and ESO269-066 that have optical colors redder than this limit (see Tab. \ref{bol}) the corrections are highly uncertain. We can however see that the resulting absolute bolometric values are consistent with the ones derived starting from the NIR. The bolometric corrections from \citet{costa96} are instead valid for the entire range in NIR colors spanned by our listed stars. We conclude that while for SGC1319.1-4216 and ESO269-066 the most luminous AGB stars have comparable bolometric luminosities, and thus similar ages (with ESO269-066 giving slightly younger values), CenA-dE1 had no significant star formation over the last $\sim9$ Gyr.

\subsection{Spatial distribution} \label{sky_sec}

\begin{figure}
 \centering
  \includegraphics[width=8cm]{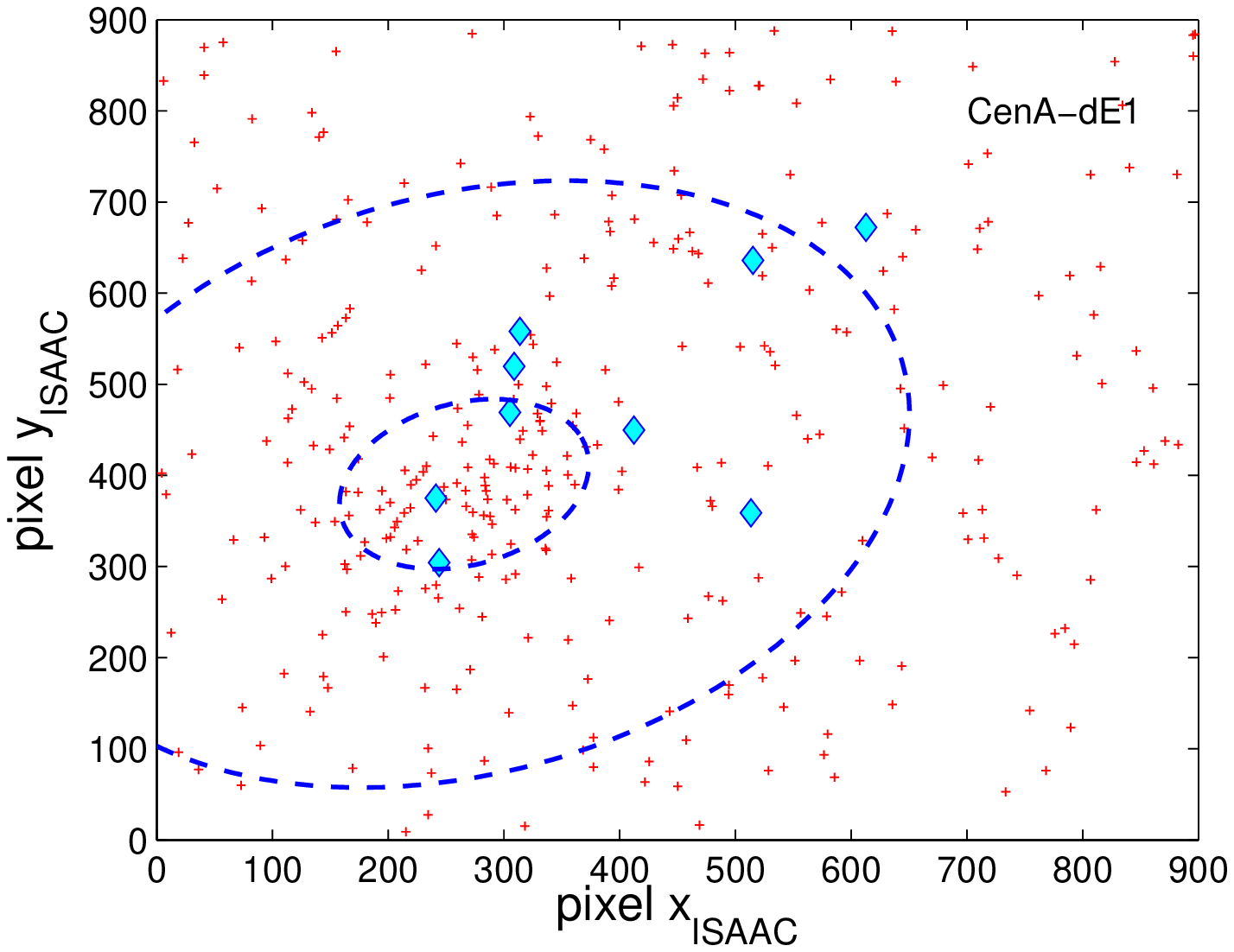}
  \includegraphics[width=8cm]{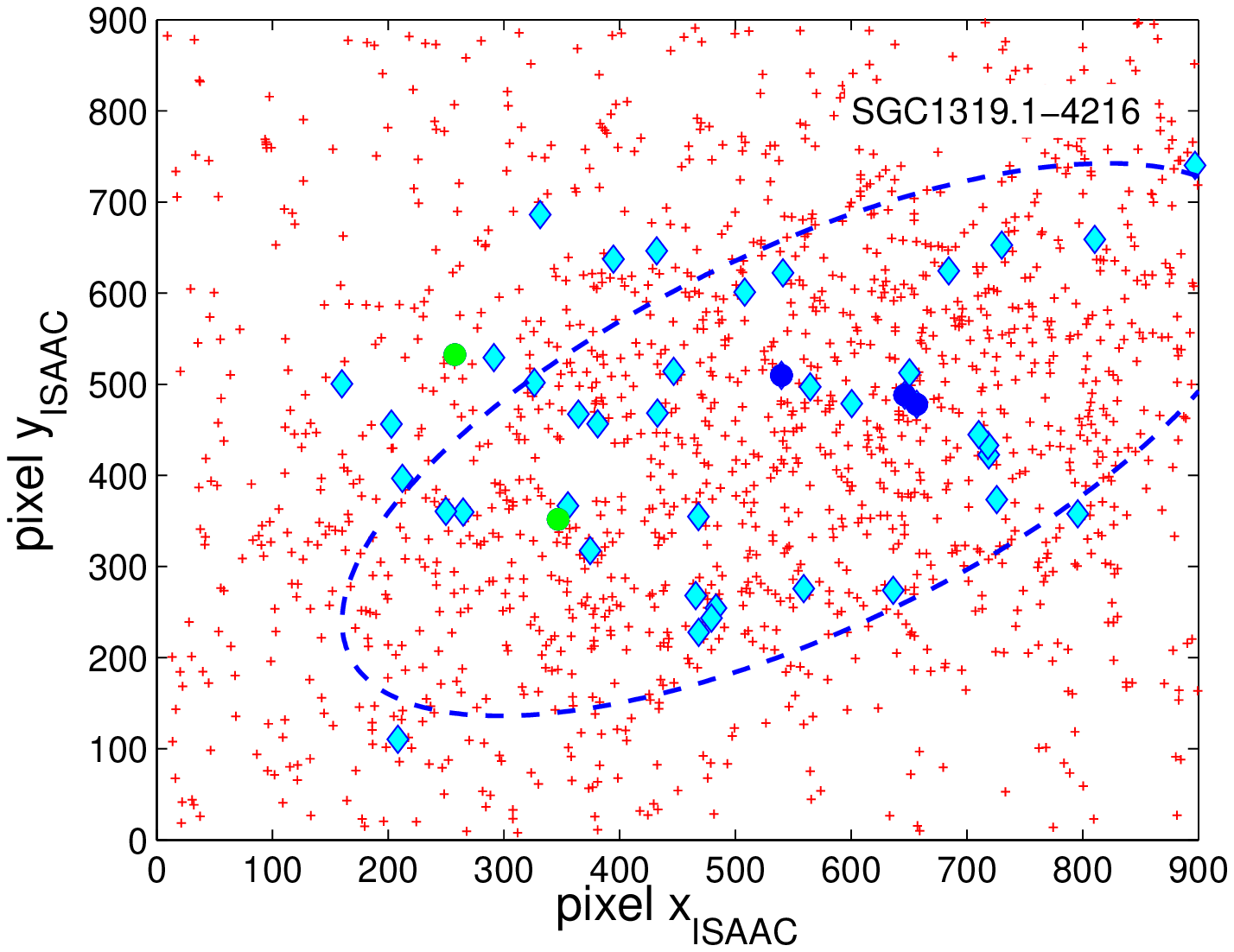}
  \includegraphics[width=8cm]{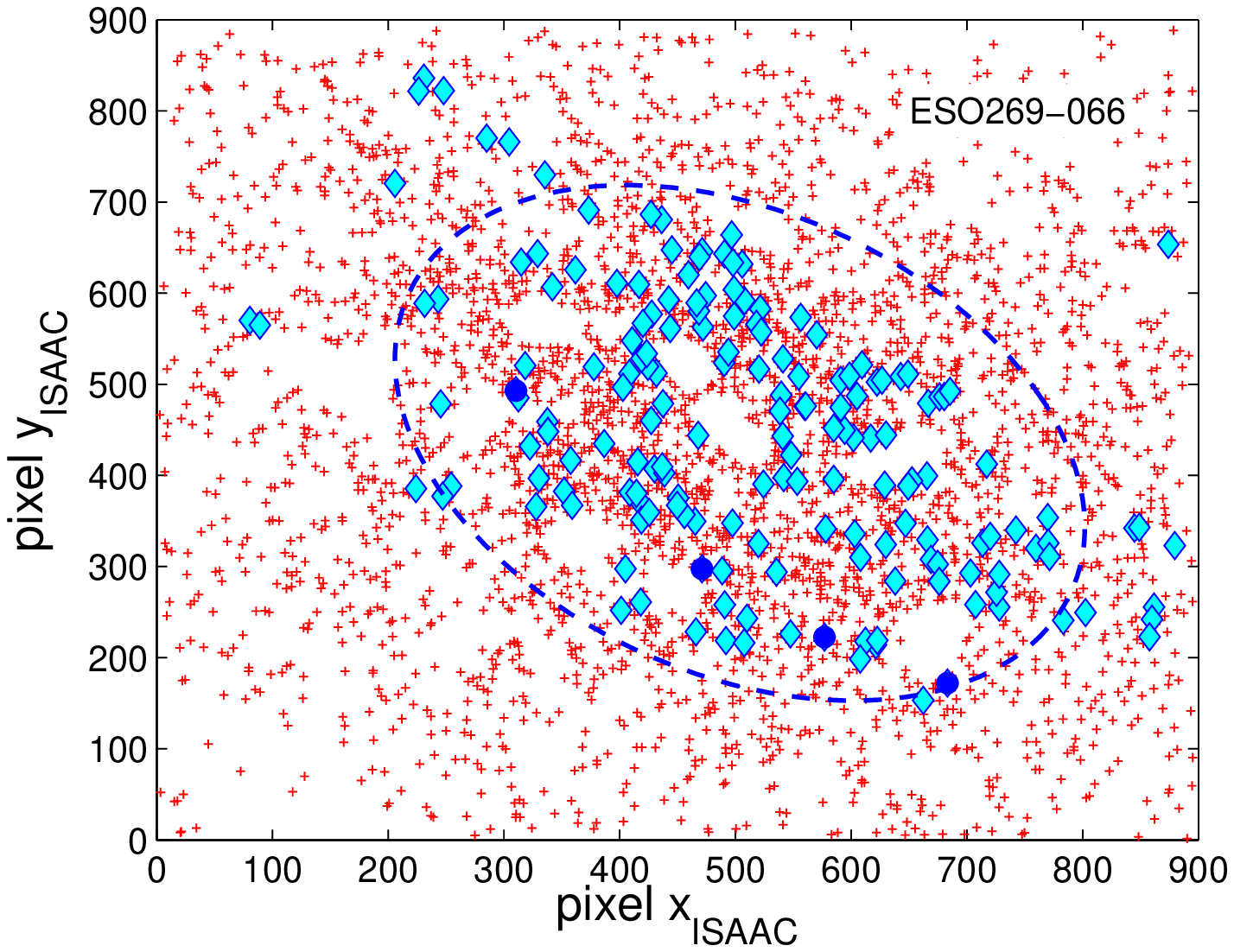}
 \caption{\footnotesize{Stellar positions projected on the sky for the target galaxies (ordered by increasing luminosity). The pixel coordinates refer to the ISAAC field-of-view. Red crosses indicate the NIR sources. The blue dashed lines are, respectively, the half-light radius and the limiting radius for CenA-dE1 and the half-light radius for SGC1319.1-4216 and ESO269-066. Cyan diamonds are candidate AGB stars, blue dots are variable stars among the candidates, and green dots are variable stars found just outside the selections boxes (see text for details). The white ``holes'' in the distribution of stars for SGC1319.1-4216 and ESO269-066 are due to bright foreground stars that are overlapping with underlying dwarf galaxy stars.}}
 \label{sky}
\end{figure}

\begin{figure}
 \centering
  \includegraphics[width=7cm]{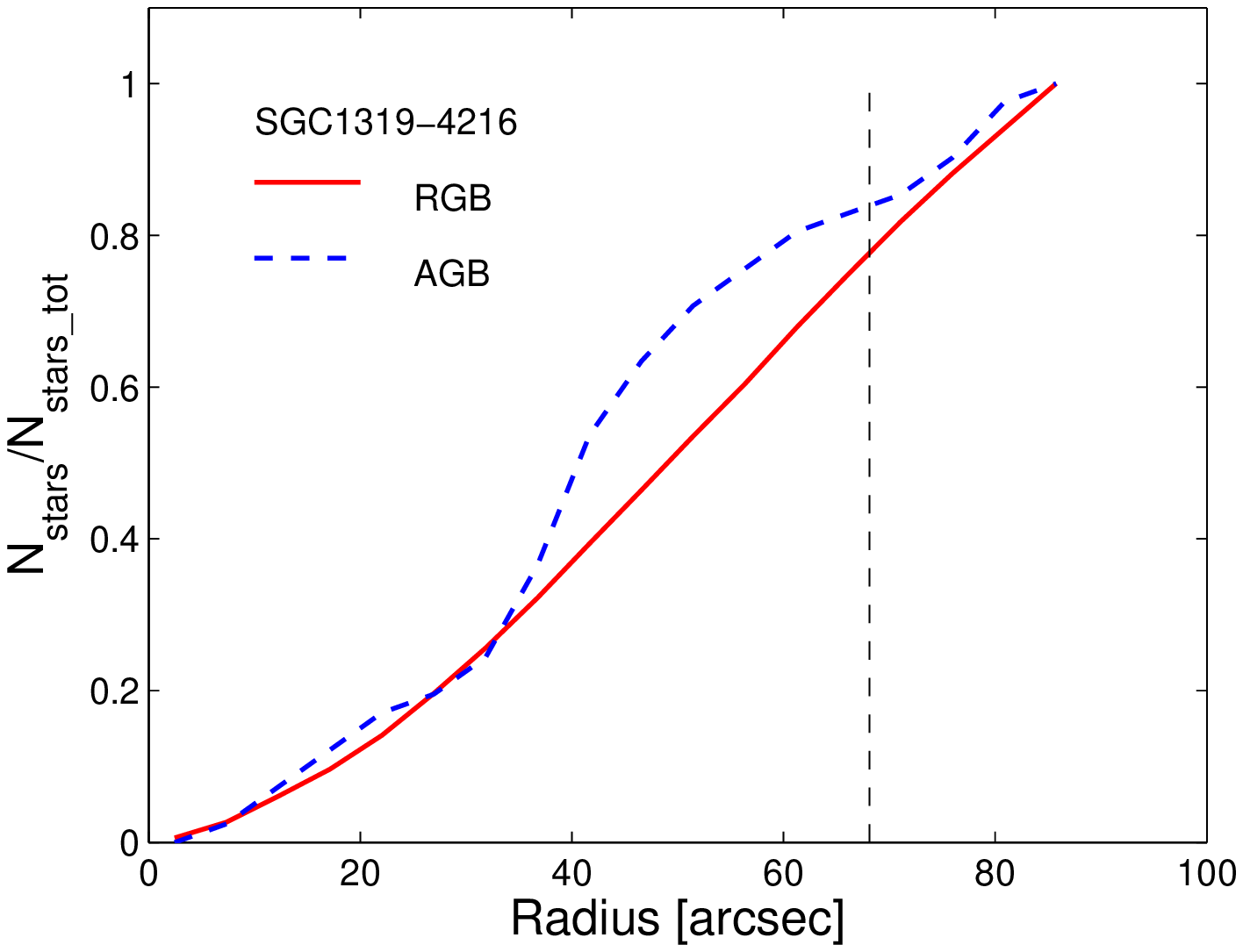}
  \includegraphics[width=7cm]{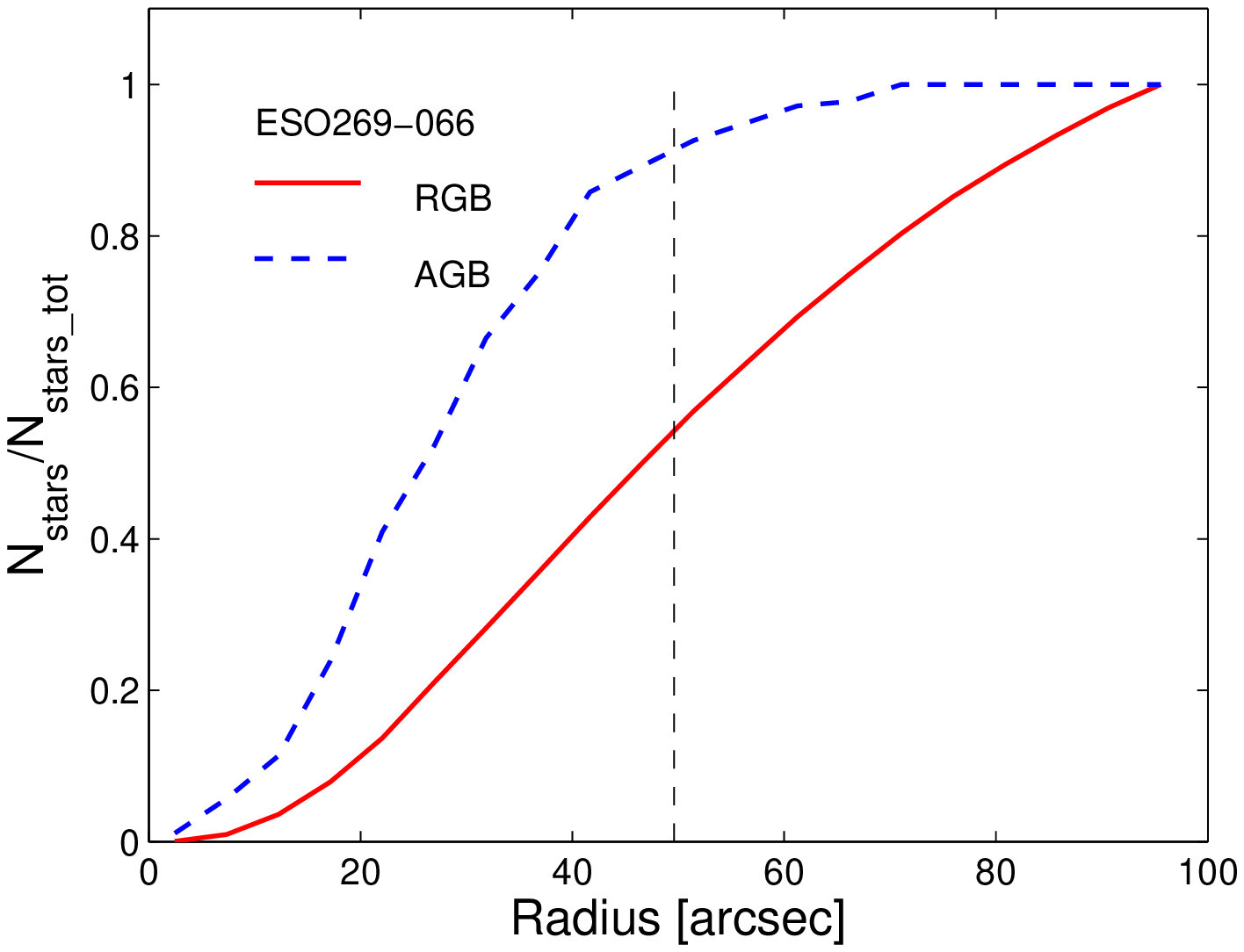}
 \caption{\footnotesize{Cumulative distribution functions of stellar number counts as a function of elliptical radius, for SGC1319.1-4216 and ESO269-066. The different curves show the distribution functions for RGB stars (red line) and candidate AGB stars (blue dashed line, see text for details about the subsamples). The vertical black dashed line indicates the half-light radius, for comparison with Fig. \ref{sky} (at these distances, 1 arcsec $\sim0.02$ kpc).}}
 \label{cumdist}
\end{figure}

We show the positions of the candidate AGB stars in Fig. \ref{sky}. The coordinate system is the one of the ISAAC instrument (in pixels), and there is complete overlap with the field of view of ACS since the latter is slightly larger. Red crosses indicate the entire sample of NIR sources (after applying the quality cuts). The AGB candidates are shown as cyan diamonds, while blue filled circles are assigned to candidate variable stars among the AGB stars and green filled circles refer to those stars that are variable but are found outside the selection boxes in the CMDs (see previous Sect.). As a reference, we overplot the half-light radius and limiting galactic radius for CenA-dE1 and the half-light radius for SGC1319.1-4216 and ESO269-066, since the last two galaxies extend beyond the field of view covered by the observations (values adopted from \citealt{crnojevic10}).

From the previous analysis of HST data \citep{crnojevic10}, we find that the ratio of AGB/RGB stars stays roughly constant as a function of radius for CenA-dE1 and SGC1319.1-4216, while for ESO269-066 there is an indication for the ratio being slightly higher in the central part of the galaxy. In Fig. \ref{sky}, we can see that in CenA-dE1 the candidate AGB stars are not symmetrically distributed, and do not tend to cluster in the center of the galaxy. This displacement is probably due to the small number statistics. In SGC1319.1-4216 the luminous AGB stars are smoothly distributed all across the galaxy, but mostly found within the half-light radius. Finally, although the AGB candidate distribution for ESO269-066 is similar to that for SGC1319.1-4216, the former has a higher density of AGB candidates in its central part. For the last two galaxies, we can notice that there are some ``holes'' in the distribution of the whole stellar sample, and these are due to bright foreground stars. In particular, given the high density of AGB candidates for ESO269-066, these foreground stars might be precluding the detection of additional AGB candidates, up to as much as $\sim10\%$ of the detected number.

We compare quantitatively the radial distribution of AGB candidates (i.e., the cyan, blue and green symbols in Fig. \ref{sky}) to that of RGB stars. For the latter, we select the stars belonging to the RGB from the HST catalog (based on their distribution in the CMD), because they have a broader spatial coverage and are more numerous than the NIR sample. For CenA-dE1 there are too few AGB candidates to draw such distributions, but we do that for SGC1319.1-4216 and ESO269-066 (Fig. \ref{cumdist}). We show the cumulative radial distribution functions (as a function of galactic elliptical radius) for both the AGB and RGB subsamples, and investigate the hypothesis that they come from the same parent distribution with a Kolmogorov-Smirnov test. For SGC1319.1-4216, there is a probability of $\sim16\%$ (i.e., less than $2\sigma$ significance level) that the two subsamples come from the same parent distribution. The difference in the cumulative distribution functions at a radius of $\sim40$ to 60 arcsec comes from the asymmetric distribution of AGB candidates, which possibly suggests an off-center star formation activity. On the other hand, for ESO269-066 we can reject the null hypothesis with a probability $<<0.1\%$, which indicates the existence of two statistically separated stellar subpopulations. Already in \citet{crnojevic10} we find that a metal-poor and a metal-rich subpopulations among the predominantly old RGB stars are clearly present for this galaxy. This result supports the evidence already found in many Local Group early-type dwarfs, for which younger and/or more metal-rich subpopulations are more centrally concentrated than the old and/or more metal-poor subpopulations \citep[e.g.][]{harbeck01}.

%________________________________________________________________

\section{Discussion} \label{discuss}

We can now compare our results to what we know for early-type dwarf galaxies in the Local Group. Among companions of the Milky Way, there are many objects that do show a substantial amount of IAPs (stars with ages in the range $\sim1-9$ Gyr), like for example the dwarf spheroidals Fornax, Carina, Leo I and Leo II (see \citealt{grebel04} and Tab. 7 of \citealt{rejkuba06}). The stellar populations of the mentioned galaxies contain as much as $\sim50\%$ of IAPs, where these estimates come from detailed SFH recovery from deep CMDs. Considering the M31 companions, the dwarf elliptical galaxies NGC205, NGC185 and NGC147 \citep[e.g.,][]{demers03, nowo03, davidge05} do show some presence of IAPs, but among the dwarf spheroidals only AndII and AndVII contain a small fraction of such young populations \citep{harbeck04, kersch04}. Generally, in dwarf galaxies with more recent star formation these populations tend to be more centrally concentrated than the older populations \citep{harbeck01}.

We mention that previous studies have only been able to put upper limits on the HI gas mass of our target galaxies. For example, \citet{bouchard07} report values of $M_{HI}/L_B<0.05$, $0.01$ and $0.002$ for CenA-dE1, SGC1319.1-4216 and ESO269-066, which are comparable to the values found for Local Group early-type dwarfs with IAPs \citep[see, e.g., Table 1 of][]{grebel03} after considering the lower HI detection limit at the distance of the CenA group. \citet{bouchard08} estimates the current rate of star formation (from H$\alpha$ emission) to be $<0.4\times10^{-5}$M$_{\odot}$yr$^{-1}$ for the target galaxies. These results all confirm their morphological classification and absence of a significant amount of ongoing star formation.

When looking at the bolometric luminosities of the brightest AGB stars of dwarfs in both the Local Group and the Centaurus A group, with our small sample we can confirm the results of \citet{rejkuba06} and conclude that they are all in the same range, with the exception of CenA-dE1 which shows a predominantly old stellar content. The physical properties (total luminosity, mean metallicity and metallicity dispersion, deprojected distance from dominant group galaxy, see Tab. \ref{infogen}) of SGC1319.1-4216 and ESO269-066 are indeed very similar to those of Fornax and Leo I, and so are also the bolometric luminosities of the brightest AGB stars \citep[e.g.,][]{demers02, menzies02, whitelock09, held10}, although we are here considering dwarf ellipticals while Fornax and Leo I are dwarf spheroidals. As pointed out by \citet{bouchard07}, the two CenA group dwarf ellipticals might have been influenced by the radio lobes of CenA, in the sense that their proximity to the giant elliptical could have played a role in the removal of their gas content. However, we stress that only the current deprojected position of these galaxies can be known, and we do not have information about their orbits so that it is difficult to draw any firm conclusion about possible effects like ram pressure stripping or tidal interactions. CenA-dE1 is on the other hand located further away from CenA ($\sim670\pm480$ kpc, see Tab. \ref{infogen}) in our sample, and is thus more similar (in distance from the dominant group galaxy, in its luminosity and metallicity content) to the most distant dwarfs of the Local Group. Its last significant star formation episode is significantly older than for the other galaxies and most probably its SFH was much less influenced by the central giant galaxy of the group. The absence of any sizeable IAP younger than 9 Gyr is somehow unexpected in a system as apparently distant from the dominant group galaxy \citep[see, e.g.,][]{bouchard08}, but on the other side Tucana and Cetus in the Local Group show similar characteristics as CenA-dE1 \citep[see, e.g.,][]{monelli10}. As a reference, we recompute the deprojected distances of the two CenA dwarf companions AM1339-445 and AM1343-452 analyzed in the previous study by \citet{rejkuba06}. We find a distance from CenA of $\sim300\pm330$ kpc and $\sim310\pm120$ kpc, respectively.

\subsection{IAP fractions}

We estimate the fraction of IAPs with respect to the total stellar content of our target galaxies in the following way: we compute the ratio of the total number of luminous AGB stars, derived in the previous Section (also including the variable stars we found outside the selection boxes), to the number of stars found in a box that extends down to 0.3 mag below the TRGB in the optical. The latter comprises both old RGB and AGB stars and also a percentage of intermediate-age RGB stars and AGB stars that are still ascending the RGB. Finally, the number of stars in the box extending below the TRGB has been subtracted for expected foreground stars, using TRILEGAL models. The ratio computed in this way is called $P_{IA}$. The values for our target galaxies are: $\sim0.07$ for CenA-dE1; $\sim0.04$ for SGC1319.1-4216; and $\sim0.11$ for ESO269-066. We notice that SGC1319.1-4216 has a particularly low $P_{IA}$, together with a very wide metallicity spread (see Tab. \ref{infogen}). SGC1319.1-4216 must have been very active at old ages, before loosing its gas reservoir and thus stopping to produce a significant amount of stars. This could be directly linked to a possible tidal interaction with CenA or ram pressure stripping by the intergalactic medium, although the orbits for our target galaxies cannot be constrained, as pointed out above.

The next step to evaluate the IAP fraction is then to consider stellar population models and compare them to our observational findings. We use the Maraston models \citep{maraston05} and follow the procedure of \citet{rejkuba06}. Namely, we first assume that there are two main subpopulations in our galaxies, one that is old (13 Gyr) and the second one with an age of 1, 2, 4, 6 or 9 Gyr (one for each different realization). This is of course a first-order simplification of the true situation, since the star formation could also have proceeded more continuously or in more than two episodes during the galaxies' lifetime. Moreover, we underline that with our data we could be missing a certain amount of IAPs and even younger stars, if we assume that the star formation rate was, in some periods, so low that an observable amount of AGB stars was simply not produced. Thus, once again, we conclude that the fractions we derive are lower limits for the true fractions of IAP. 

The model we adopt has a metallicity of [Z/H]$=-1.35$, which matches well with all of the galaxies in our sample, if we consider that the lifetimes of the AGB phase do not dramatically change as a function of the small differences in metallicity among our target galaxies. We then let the IAP fraction vary among the different realizations (from 0 to 1), and each time compute the $P_{IA}$ ratio given by the models. When we compare the results to the observed values found above, we find that CenA-dE1 has an IAP fraction of $\sim15\%$; for SGC1319.1-4216 we get an upper limit of only $\sim5\%$; while for ESO269-066 the value we find is between $\sim5\%$ and $\sim10\%$. While the empirical relation between age and bolometric luminosity at the AGB tip that we adopt in Sect. \ref{age_sec} comes from SMC and LMC clusters \citep{rejkuba06}, the metallicity dependence of the luminosity is smaller than our errors on the bolometric magnitudes themselves. The derived ages of our AGB candidates are thus not significantly metallicity dependent, and the assumption of a metallicity [Z/H]$=-1.35$ for our model will not introduce a bias in our results.

Compared to the rough predictions given in \citet{crnojevic10} (see Sect. \ref{cmdopt_sec}), the current results are slightly higher for CenA-dE1, and lower for both SGC1319.1-4216 and ESO269-066. We check whether this would change the results about the metallicity content of CenA-dE1, and find that the median metallicity value does not change, while the metallicity spread is smaller by a few percent. For the other two galaxies there are no significant changes. We emphasize though that recent studies have pointed out how stellar evolutionary models tend to overpredict the number of luminous AGB stars for dwarfs with low metallicities, similar to our targets, by a factor of as much as $\sim3-6$ \citep[e.g.,][]{gullieu08, melbourne10}. This, together with the arguments mentioned above, reinforces the idea that our results are lower limits for the true IAP fractions in these galaxies.

Having computed the IAP fractions, even considering that the real fractions could be twice or three times those derived here, we still can clearly see that they are much lower than the fractions found in similar objects of the Local Group. We underline that the relative differences with the Local Group will not be affected by the model uncertainties, since the comparison is observationally based, although the absolute values of both sets might be larger. In particular, we mentioned that the characteristics of SGC1319.1-4216 and ESO269-066 are similar to those of Fornax and Leo I, but despite this similarity, the recent star formation has proceeded at a much lower rate for the (so far studied) objects in the Centaurus A group. The latter, denser, environment could thus be responsible for the observed properties of its dwarf members. These results are again in agreement to what is found by \citet{rejkuba06}, who also found quite low IAP fractions in two dwarf members of this group that resemble the low luminosity companions of M31. We however stress that currently the sample is too small to draw firmer conclusions, and we would need to have NIR data also for the other CenA dwarfs with existing optical HST data.

%________________________________________________________________

\section{Conclusions} \label{conclus}

We study archival photometric data for three early-type dwarf galaxies in the Centaurus A group, the dwarf spheroidal CenA-dE1 and the dwarf ellipticals SGC1319.1-4216 and ESO269-066. We select the targets based on the availability of both high-resolution HST/ACS data and ground-based VLT/ISAAC data, in order to analyze the luminous AGB stars in these objects. These stars are useful tracers of IAPs ($\sim1$ to 9 Gyr). From the optical data alone AGB stars are difficult to separate from the Galactic foreground contamination, by which the Centaurus A group region is affected.

Selecting common stars from the combined optical and NIR CMDs, we isolate 9 candidate luminous AGB stars for CenA-dE1, 41 for SGC1319.1-4216 and 176 for ESO269-066. To these, we can add 2 more stars in SGC1319.1-4216, based on their variability between the sets of $K$-band observations. We then compute the absolute bolometric magnitudes for the most luminous AGB candidates, and find that they are lower for CenA-dE1, while SGC1319.1-4216 and ESO269-066 have comparable values. We further correlate the absolute bolometric magnitudes to the age of the most recent significant star formation episode in these galaxies. We find a value of $9.0\pm1.0$ Gyr for CenA-dE1, $\sim2.0\pm1.5$ to $\sim5.0\pm2.0$ for SGC1319.1-4216, and $2.0\pm1.5$ to $3.0\pm1.5$ Gyr for ESO269-066. The two values for SGC1319.1-4216 and ESO269-066 correspond to the different methods of computing bolometric magnitudes, starting either from optical or from NIR data, while for CenA-dE1 the two methods give almost identical results. 

We find a quantitative evidence for the presence of two separated subpopulations, namely AGB and RGB stars (i.e., intermediate-age and old subpopulations) for ESO269-066. The AGB subpopulation appears to be more centrally concentrated than the RGB stars, thus resembling similar trends found in Local Group early-type dwarfs. Finally, we estimate the fractions of IAPs of the three target galaxies starting from stellar evolutionary models, and find values of $\sim15\%$ for CenA-dE1, an upper limit of only $\sim5\%$ for SGC1319.1-4216, and a value between $\sim5\%$ and $\sim10\%$ for ESO269-066. These are likely to be lower limits to the true fractions.

When comparing our results to dwarf galaxies of the Local Group, we clearly see that despite having similar physical properties and positions within the two groups, the objects in the Centaurus A group tend to have a lower IAP fraction than observed in the dwarf spheroidal companions of similar luminosity of the Milky Way. Our results are comparable to those found for low-mass dwarf spheroidal companions of M31. We suggest that this difference might be due to environmental effects, which are leading to the loss of the neutral gas content for the dwarfs in this dense group, and preferentially for objects that are closest to the dominant elliptical CenA. However, we stress that our sample, together with the study of \citet{rejkuba06}, consists of only five galaxies, so these are just preliminary findings. Finally, we underline that this kind of study has just begun for dwarf galaxies outside the Local Group and that there is need for more observational data.

%________________________________________________________________

\begin{acknowledgements}

We thank an anonymous referee who helped to improve the paper. We thank the service mode support at Paranal for conducting the near-infrared observations. DC acknowledges financial support from the MPIA of Heidelberg, as part of the IMPRS program, and from the Astronomisches Rechen-Institut of the University of Heidelberg. DC is thankful to S. Pasetto and S. Jin for enlightening conversations and support. This publication made use of data products from the Two Micron All Sky Survey, which is a joint project of the University of Massachussetts and the Infrared Processing and Analysis Center/California Institute of Technology, funded by the National Science Foundation. This work is also based on observations made with the NASA/ESA Hubble Space Telescope, obtained from the data archive at the Space Telescope Science Institute. STScI is operated by the Association of Universities for Research in Astronomy, Inc. under NASA contract NAS 5-26555. This research made use of the NASA/IPAC Extragalactic Database (NED), which is operated by the Jet Propulsion Laboratory, California Institute of Technology, under contract with the National Aeronautics and Space Administration.

\end{acknowledgements}

%________________________________________________________________

\bibliographystyle{aa}
\bibliography{biblio.bib}

\begin{thebibliography}{54}
\expandafter\ifx\csname natexlab\endcsname\relax\def\natexlab#1{#1}\fi

\bibitem[{{Armandroff} {et~al.}(1993){Armandroff}, {Da Costa}, {Caldwell}, \&
  {Seitzer}}]{armand93}
{Armandroff}, T.~E., {Da Costa}, G.~S., {Caldwell}, N., \& {Seitzer}, P. 1993,
  \aj, 106, 986

\bibitem[{{Banks} {et~al.}(1999){Banks}, {Disney}, {Knezek}, {Jerjen},
  {Barnes}, {Bhatal}, {de Blok}, {Boyce}, \& {et al.}}]{banks99}
{Banks}, G.~D., {Disney}, M.~J., {Knezek}, P.~M., {et~al.} 1999, \apj, 524, 612

\bibitem[{{Bouchard} {et~al.}(2009){Bouchard}, {Da Costa}, \&
  {Jerjen}}]{bouchard08}
{Bouchard}, A., {Da Costa}, G.~S., \& {Jerjen}, H. 2009, \aj, 137, 3038

\bibitem[{{Bouchard} {et~al.}(2007){Bouchard}, {Jerjen}, {Da Costa}, \&
  {Ott}}]{bouchard07}
{Bouchard}, A., {Jerjen}, H., {Da Costa}, G.~S., \& {Ott}, J. 2007, \aj, 133,
  261

\bibitem[{{Boyer} {et~al.}(2009){Boyer}, {Skillman}, {van Loon}, {Gehrz}, \&
  {Woodward}}]{boyer09}
{Boyer}, M.~L., {Skillman}, E.~D., {van Loon}, J.~T., {Gehrz}, R.~D., \&
  {Woodward}, C.~E. 2009, \apj, 697, 1993

\bibitem[{{Carpenter}(2001)}]{carpenter01}
{Carpenter}, J.~M. 2001, \aj, 121, 2851

\bibitem[{{Costa} \& {Frogel}(1996)}]{costa96}
{Costa}, E. \& {Frogel}, J.~A. 1996, \aj, 112, 2607

\bibitem[{{C{\^o}t{\'e}} {et~al.}(1997){C{\^o}t{\'e}}, {Freeman}, {Carignan},
  \& {Quinn}}]{cote97}
{C{\^o}t{\'e}}, S., {Freeman}, K.~C., {Carignan}, C., \& {Quinn}, P.~J. 1997,
  \aj, 114, 1313

\bibitem[{{Crnojevi{\'c}} {et~al.}(2010){Crnojevi{\'c}}, {Grebel}, \&
  {Koch}}]{crnojevic10}
{Crnojevi{\'c}}, D., {Grebel}, E.~K., \& {Koch}, A. 2010, \aap, 516, A85

\bibitem[{{Da Costa} \& {Armandroff}(1990)}]{dacosta90}
{Da Costa}, G.~S. \& {Armandroff}, T.~E. 1990, \aj, 100, 162

\bibitem[{{Davidge}(2005)}]{davidge05}
{Davidge}, T.~J. 2005, \aj, 130, 2087

\bibitem[{{Demers} {et~al.}(2003){Demers}, {Battinelli}, \&
  {Letarte}}]{demers03}
{Demers}, S., {Battinelli}, P., \& {Letarte}, B. 2003, \aj, 125, 3037

\bibitem[{{Demers} {et~al.}(2002){Demers}, {Dallaire}, \&
  {Battinelli}}]{demers02}
{Demers}, S., {Dallaire}, M., \& {Battinelli}, P. 2002, \aj, 123, 3428

\bibitem[{{Dolphin}(2002)}]{dolphin02}
{Dolphin}, A.~E. 2002, \mnras, 332, 91

\bibitem[{{Dotter} {et~al.}(2008){Dotter}, {Chaboyer}, {Jevremovi{\'c}},
  {Kostov}, {Baron}, \& {Ferguson}}]{dotter08}
{Dotter}, A., {Chaboyer}, B., {Jevremovi{\'c}}, D., {et~al.} 2008, \apjs, 178,
  89

\bibitem[{{Fleming} {et~al.}(1995){Fleming}, {Harris}, {Pritchet}, \&
  {Hanes}}]{fleming95}
{Fleming}, D.~E.~B., {Harris}, W.~E., {Pritchet}, C.~J., \& {Hanes}, D.~A.
  1995, \aj, 109, 1044

\bibitem[{{Georgiev} {et~al.}(2008){Georgiev}, {Goudfrooij}, {Puzia}, \&
  {Hilker}}]{georgiev08}
{Georgiev}, I.~Y., {Goudfrooij}, P., {Puzia}, T.~H., \& {Hilker}, M. 2008, \aj,
  135, 1858

\bibitem[{{Girardi} {et~al.}(2005){Girardi}, {Groenewegen}, {Hatziminaoglou},
  \& {da Costa}}]{girardi05}
{Girardi}, L., {Groenewegen}, M.~A.~T., {Hatziminaoglou}, E., \& {da Costa}, L.
  2005, \aap, 436, 895

\bibitem[{{Girardi} {et~al.}(2010){Girardi}, {Williams}, {Gilbert},
  {Rosenfield}, {Dalcanton}, {Marigo}, {Boyer}, {Dolphin}, {Weisz},
  {Melbourne}, {Olsen}, {Seth}, \& {Skillman}}]{girardi10}
{Girardi}, L., {Williams}, B.~F., {Gilbert}, K.~M., {et~al.} 2010,
  ArXiv:1009.4618

\bibitem[{{Grebel}(1997)}]{grebel97}
{Grebel}, E.~K. 1997, in Reviews in Modern Astronomy, ed. R.~E. {Schielicke},
  Vol.~10, 29

\bibitem[{{Grebel} \& {Gallagher}(2004)}]{grebel04}
{Grebel}, E.~K. \& {Gallagher}, III, J.~S. 2004, \apjl, 610, L89

\bibitem[{{Grebel} {et~al.}(2003){Grebel}, {Gallagher}, \&
  {Harbeck}}]{grebel03}
{Grebel}, E.~K., {Gallagher}, III, J.~S., \& {Harbeck}, D. 2003, \aj, 125, 1926

\bibitem[{{Gullieuszik} {et~al.}(2008){Gullieuszik}, {Held}, {Rizzi},
  {Girardi}, {Marigo}, \& {Momany}}]{gullieu08}
{Gullieuszik}, M., {Held}, E.~V., {Rizzi}, L., {et~al.} 2008, \mnras, 388, 1185

\bibitem[{{Harbeck} {et~al.}(2004){Harbeck}, {Gallagher}, \&
  {Grebel}}]{harbeck04}
{Harbeck}, D., {Gallagher}, III, J.~S., \& {Grebel}, E.~K. 2004, \aj, 127, 2711

\bibitem[{{Harbeck} {et~al.}(2001){Harbeck}, {Grebel}, {Holtzman},
  {Guhathakurta}, {Brandner}, {Geisler}, {Sarajedini}, {Dolphin}, \& {et
  al.}}]{harbeck01}
{Harbeck}, D., {Grebel}, E.~K., {Holtzman}, J., {et~al.} 2001, \aj, 122, 3092

\bibitem[{{Harris} {et~al.}(2009){Harris}, {Rejkuba}, \& {Harris}}]{harrisg09}
{Harris}, G.~L.~H., {Rejkuba}, M., \& {Harris}, W.~E. 2009, ArXiv:0911.3180

\bibitem[{{Held} {et~al.}(2010){Held}, {Gullieuszik}, {Rizzi}, {Girardi},
  {Marigo}, \& {Saviane}}]{held10}
{Held}, E.~V., {Gullieuszik}, M., {Rizzi}, L., {et~al.} 2010, \mnras, 404, 1475

\bibitem[{{Jerjen} {et~al.}(2000){Jerjen}, {Binggeli}, \&
  {Freeman}}]{jerjen00b}
{Jerjen}, H., {Binggeli}, B., \& {Freeman}, K.~C. 2000, \aj, 119, 593

\bibitem[{{Kang} {et~al.}(2006){Kang}, {Sohn}, {Kim}, {Rhee}, {Kim}, {Hwang},
  {Lee}, {Kim}, \& {et al.}}]{kang06}
{Kang}, A., {Sohn}, Y., {Kim}, H., {et~al.} 2006, \aap, 454, 717

\bibitem[{{Karachentsev}(2005)}]{kara05}
{Karachentsev}, I.~D. 2005, \aj, 129, 178

\bibitem[{{Karachentsev} {et~al.}(2002){Karachentsev}, {Sharina}, {Dolphin},
  {Grebel}, {Geisler}, {Guhathakurta}, {Hodge}, {Karachentseva}, \& {et
  al.}}]{kara02}
{Karachentsev}, I.~D., {Sharina}, M.~E., {Dolphin}, A.~E., {et~al.} 2002, \aap,
  385, 21

\bibitem[{{Karachentsev} {et~al.}(2007){Karachentsev}, {Tully}, {Dolphin},
  {Sharina}, {Makarova}, {Makarov}, {Sakai}, {Shaya}, \& {et al.}}]{kara07}
{Karachentsev}, I.~D., {Tully}, R.~B., {Dolphin}, A., {et~al.} 2007, \aj, 133,
  504

\bibitem[{{Kerschbaum} {et~al.}(2004){Kerschbaum}, {Nowotny}, {Olofsson}, \&
  {Schwarz}}]{kersch04}
{Kerschbaum}, F., {Nowotny}, W., {Olofsson}, H., \& {Schwarz}, H.~E. 2004,
  \aap, 427, 613

\bibitem[{{Lee} {et~al.}(1993){Lee}, {Freedman}, \& {Madore}}]{lee93}
{Lee}, M.~G., {Freedman}, W.~L., \& {Madore}, B.~F. 1993, \apj, 417, 553

\bibitem[{{Lianou} {et~al.}(2010){Lianou}, {Grebel}, \& {Koch}}]{lianou10}
{Lianou}, S., {Grebel}, E.~K., \& {Koch}, A. 2010, ArXiv:1003.0861

\bibitem[{{Maraston}(2005)}]{maraston05}
{Maraston}, C. 2005, \mnras, 362, 799

\bibitem[{{Melbourne} {et~al.}(2010){Melbourne}, {Williams}, {Dalcanton},
  {Ammons}, {Max}, {Koo}, {Girardi}, \& {Dolphin}}]{melbourne10}
{Melbourne}, J., {Williams}, B., {Dalcanton}, J., {et~al.} 2010, \apj, 712, 469

\bibitem[{{Menzies} {et~al.}(2002){Menzies}, {Feast}, {Tanab{\'e}},
  {Whitelock}, \& {Nakada}}]{menzies02}
{Menzies}, J., {Feast}, M., {Tanab{\'e}}, T., {Whitelock}, P., \& {Nakada}, Y.
  2002, \mnras, 335, 923

\bibitem[{{Monelli} {et~al.}(2010){Monelli}, {Hidalgo}, {Stetson}, {Aparicio},
  {Gallart}, {Dolphin}, {Cole}, {Weisz}, \& {et al.}}]{monelli10}
{Monelli}, M., {Hidalgo}, S.~L., {Stetson}, P.~B., {et~al.} 2010, \apj, 720,
  1225

\bibitem[{{Nowotny} {et~al.}(2003){Nowotny}, {Kerschbaum}, {Olofsson}, \&
  {Schwarz}}]{nowo03}
{Nowotny}, W., {Kerschbaum}, F., {Olofsson}, H., \& {Schwarz}, H.~E. 2003,
  \aap, 403, 93

\bibitem[{{Puzia} {et~al.}(1999){Puzia}, {Kissler-Patig}, {Brodie}, \&
  {Huchra}}]{puzia99}
{Puzia}, T.~H., {Kissler-Patig}, M., {Brodie}, J.~P., \& {Huchra}, J.~P. 1999,
  \aj, 118, 2734

\bibitem[{{Rejkuba} {et~al.}(2006){Rejkuba}, {Da Costa}, {Jerjen}, {Zoccali},
  \& {Binggeli}}]{rejkuba06}
{Rejkuba}, M., {Da Costa}, G.~S., {Jerjen}, H., {Zoccali}, M., \& {Binggeli},
  B. 2006, \aap, 448, 983

\bibitem[{{Rejkuba} {et~al.}(2001){Rejkuba}, {Minniti}, {Silva}, \&
  {Bedding}}]{rejkuba01}
{Rejkuba}, M., {Minniti}, D., {Silva}, D.~R., \& {Bedding}, T.~R. 2001, \aap,
  379, 781

\bibitem[{{Rejkuba} {et~al.}(2003){Rejkuba}, {Minniti}, {Silva}, \&
  {Bedding}}]{rejkuba03}
{Rejkuba}, M., {Minniti}, D., {Silva}, D.~R., \& {Bedding}, T.~R. 2003, \aap,
  411, 351

\bibitem[{{Renzini} \& {Buzzoni}(1986)}]{renzini86}
{Renzini}, A. \& {Buzzoni}, A. 1986, in Astrophysics and Space Science Library,
  Vol. 122, Spectral Evolution of Galaxies, ed. {C.~Chiosi \& A.~Renzini}, 195

\bibitem[{{Robin} {et~al.}(2003){Robin}, {Reyl{\'e}}, {Derri{\`e}re}, \&
  {Picaud}}]{robin03}
{Robin}, A.~C., {Reyl{\'e}}, C., {Derri{\`e}re}, S., \& {Picaud}, S. 2003,
  \aap, 409, 523

\bibitem[{{Saracco} {et~al.}(2001){Saracco}, {Giallongo}, {Cristiani},
  {D'Odorico}, {Fontana}, {Iovino}, {Poli}, \& {Vanzella}}]{saracco01}
{Saracco}, P., {Giallongo}, E., {Cristiani}, S., {et~al.} 2001, \aap, 375, 1

\bibitem[{{Schlegel} {et~al.}(1998){Schlegel}, {Finkbeiner}, \&
  {Davis}}]{schlegel98}
{Schlegel}, D.~J., {Finkbeiner}, D.~P., \& {Davis}, M. 1998, \apj, 500, 525

\bibitem[{{Stetson}(1987)}]{stetson87}
{Stetson}, P.~B. 1987, \pasp, 99, 191

\bibitem[{{Stetson}(1994)}]{stetson94}
{Stetson}, P.~B. 1994, \pasp, 106, 250

\bibitem[{{Tolstoy} {et~al.}(2009){Tolstoy}, {Hill}, \& {Tosi}}]{tolstoy09}
{Tolstoy}, E., {Hill}, V., \& {Tosi}, M. 2009, \araa, 47, 371

\bibitem[{{Valenti} {et~al.}(2004){Valenti}, {Ferraro}, \&
  {Origlia}}]{valenti04}
{Valenti}, E., {Ferraro}, F.~R., \& {Origlia}, L. 2004, \mnras, 354, 815

\bibitem[{{van Loon} {et~al.}(2005){van Loon}, {Marshall}, \&
  {Zijlstra}}]{vanloon05}
{van Loon}, J.~T., {Marshall}, J.~R., \& {Zijlstra}, A.~A. 2005, \aap, 442, 597

\bibitem[{{Whitelock} {et~al.}(2009){Whitelock}, {Menzies}, {Feast},
  {Matsunaga}, {Tanab{\'e}}, \& {Ita}}]{whitelock09}
{Whitelock}, P.~A., {Menzies}, J.~W., {Feast}, M.~W., {et~al.} 2009, \mnras,
  394, 795

\end{thebibliography}

%________________________________________________________________

\end{document}